\documentclass[10pt,aps,prx,longbibliography,]{revtex4-2}

\usepackage[utf8]{inputenc}
\usepackage{graphicx} 
\usepackage{amsmath,amssymb,amsfonts,bm}
\usepackage{bbm}
\usepackage{hyperref}
\usepackage{cleveref}

\begin{document}
\title{Random quantum circuits, chaos and quantum thermalization}
\author{J. T. Chalker}
\address{Rudolf Peierls Centre for Theoretical Physics, Physics Department, Oxford University, Parks Road, Oxford OX1 3PU, United Kingdom}
\date{\today}


\begin{abstract}
These notes accompany lectures given in June 2025 at the summer school \emph{Fundamental Problems in Statistical Physics XVI}. 
They offer a short introduction to random quantum circuits as simple models for generic many-body quantum systems. They give an outline of the motivation for introducing these models, starting from ideas of random matrix theory. They also provide a sketch of calculations of some of the quantities of most physical interest, based on an average over an ensemble of systems. These quantities give insights into  operator spreading, entanglement dynamics and spectral correlations.
\end{abstract}

\maketitle

\section{Introduction}

These notes accompany lectures given in June 2025 at the summer school \emph{Fundamental Problems in Statistical Physics XVI}. They are intended to provide a short introduction to random quantum circuits as simple models for quantum dynamics in generic, spatially extended systems, and to outline what has been learned from studying these models. Some review articles covering this material in more detail are Refs.~\cite{Potter_2022} and \cite{Fisher_Random_2023}. We make only a very limited number of references to the literature here, and much more extensive citations can be found in these reviews. Background on chaotic quantum systems and random matrix theory is described in \cite{DAlessio_2016} and \cite{Haake_2010}, and the fundamentals of random matrix theory are set out in \cite{Mehta_Random_2004}. These notes and lectures do not touch on the experimental platforms being developed to probe some of the phenomena we discuss; for an introduction to that area, see the contributions to this school from Marcello Dalmonte. 

To set the context, it is useful to start by considering many-body quantum physics from a general perspective, and to identify the features that distinguish the problems we discuss here. As a first example, take the ideal Bose and Fermi gases, which for most physicists represent their first encounter with many-body quantum systems. The ideal gases are simple to treat because they have an extensive number of conserved quantities, which are the particle numbers in each single-particle orbital. This feature of an extensive number of conserved quantities is of course very special, although it is shared by one-dimensional interacting systems that have Hamiltonians specially chosen to be integrable using the Bethe Ansatz. By contrast, our focus will be on systems that have a small, fixed number of local conserved densities, independent of system size, or possibly none at all. 

A second important class of many-body problem arises at low temperature for systems that have quasiparticle description based on long-lived excitations which share some of the properties of an ideal quantum gas. Examples include interacting Fermi and Bose liquids, and ordered magnets. By contrast again, our focus will be on systems in which there are no long-lived quasiparticles. We are hence concerned with interacting systems that are generic, rather than fine-tuned, and with dynamics in generic states, rather than close to the ground state. 

Continuing with our discussion of the broad context, we can note that physicists in the past have been very successful at developing approaches tailored to different types of problem: for example, many-body perturbation theory as a framework for quasiparticle descriptions, or trial wavefunctions as instances of topologically ordered states. Against that background, random circuits represent an approach to treating generic many-body systems, with an emphasis on the consequences of local interactions in spatially extended models. Since, as we will see, these models are highly stylised, it makes sense to use them only to calculate features that show a degree of universality, in the sense of being independent of model details. This is most likely to be the case if we probe evolution of a system at long times compared to the inverse of the coupling energies, and long distances compared to the lattice spacing. Equally, it is reasonable to look for universal features in the spectral properties of the evolution operator, provided we focus on scales much finer than the one set by interaction strength. 

\section{What do we want to know?}

The behaviour of many-body quantum systems close to equilibrium is well characterised by time-ordered correlation functions. In particular, these describe the response of a system to the established experimental probes used in condensed matter physics. In the regime we are concerned with, far from the ground state, many such correlations decay rapidly. The exceptions are ones involving conserved densities: the dynamics of these is expected to be described by hydrodynamic theories such as diffusion, which can arise equally in classical systems. Information about the underlying quantum system then appears at long times and distances only via the values of transport coefficients.

Here, by contrast, our interest will be mainly in probes of coherent many-body dynamics that characterise the spread of quantum information, or that are recognised as an indicator of chaos. One focus is to understand how unitary time evolution, despite being in principle reversible and without loss of information, may nevertheless lead to an apparent equilibrium state. We will treat three separate phenomena. First, considering time evolution in the Heisenberg picture, one asks about the evolution of operators. In general, one expects that an operator that is initially simple will evolve into something increasingly complicated. More specifically, an operator that acts locally on the system will evolve into an operator that acts on a number of sites that increases with time: we will study this operator growth in detail for random quantum circuits. Second, considering time evolution in the Schr\"odinger picture, one asks about the evolution of state vectors and the density matrix. It is often natural to think about evolution from an initial state that is simple, in the sense that it is a product state in a site basis or at least has low entanglement across spatial cuts. Then we expect time evolution to take this state to one that is more complicated, in the sense that it has higher entanglement across spatial cuts. We will study this process via the behaviour of the density matrix in random quantum circuits. Third, one can consider properties of the evolution operator itself, of which the simplest are the spectral correlations that are the central object of study in random matrix theory. Random matrix spectral correlations on appropriate scales are believed to be an indicator of quantum chaos, and here our concern is both with how these arise and with the new features that may appear in a locally coupled, spatially extended system.    

\section{How can we do calculations?}

Our next step is to set up simple models for generic spatially extended many-body quantum systems. Important inspiration comes from work on random matrix theory done in the 1950s and 1960s. A question at that time was how to think about the properties of sequences of a few tens of high-lying nuclear energy energy levels, measured in scattering experiments (see \cite{Brody_1981} for a review). Having in mind the successes of the theory of electronic levels in simple atoms, a possible approach might be to try to calculate these levels starting from a microscopic model of the nucleus. That, however, is not appropriate, for multiple reasons. Practically, it is challenging because of the sizes of the relevant Hilbert spaces, but more importantly, the results are expected to be highly sensitive to details of the model, which are not known with sufficient accuracy. The latter is not just a practical difficulty but also a conceptual one, since it implies that the interesting information does not reside in the values of individual energy levels. 

A revolutionary change in viewpoint is to think instead about statistical properties of the sequence of levels. In Dyson's words \cite{Dyson_1962}, we should \emph{renounce knowledge of $\ldots$ the nature of the system}. We can do this by studying an ensemble of Hamiltonians instead of a single instance. Crucially, while properties of a specific generic Hamiltonian are (essentially on principle) accessible only numerically, the same properties averaged over an ensemble of Hamiltonians may be reasonably simple to evaluate analytically if the ensemble is chosen to facilitate calculations. We can think of the ensemble as representing an average over a sufficiently long sequence of levels, or as an average over properties of a number of distinct but macroscopically similar systems.

While random matrix theory in this context was first developed for Hermitian matrices as models for the Hamiltonian of a system, we will be concerned with time evolution operators, viewed as unitary matrices acting on a Hilbert space of dimension $N$. Since the space U(N) is compact, it is straightforward to define a uniform probability distribution on it, which is known as the Haar distribution. The resulting matrix ensemble is also called the circular unitary ensemble (CUE), since all eigenvalues lie on the unit circle in the complex plane \cite{Dyson_1962}. Formally, the Haar distribution is the distribution that is invariant under multiplication of all members by the same fixed but arbitrary unitary matrix, either from the left or from the right. As we will see, averages over this ensemble of products of matrix elements are easy to evaluate at low order, but become increasingly difficult at higher order, unless $N$ is large. 

We are now in a position to introduce random quantum circuits. The first ingredient, already discussed, is the idea that we should consider ensembles of systems rather than individual instances. The second, which is less obvious, is that it is much simpler to model the evolution operator directly, instead of first defining an ensemble of Hamiltonians and using this to construct the evolution operator (recent progress via the second route is described in \cite{Chalker_2025}). The third ingredient is locality, meaning that evolution over short times should couple only nearby parts of the system. We apply these ingredients to a generalised spin system, in which each site of a lattice caries a local Hilbert space of dimension $q$, with their tensor product giving the Hilbert space for the entire system; in the simplest case of spin-half we have $q=2$. We will discuss only systems in one space dimension, so that the sites form a chain, although this restriction is not fundamental. 

\begin{figure}[htb]
    \centering 
      $|\psi \rangle \equiv$  \includegraphics[width=1cm]{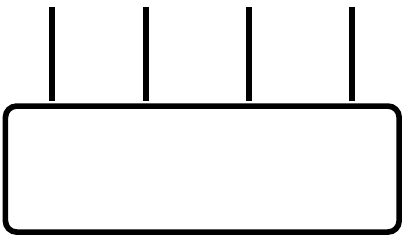}\,,\hspace{0.5cm}
      $\langle  \uparrow\uparrow\downarrow \downarrow |\psi\rangle \equiv $ \includegraphics[width=1cm]{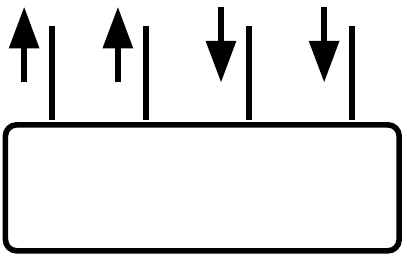} \hspace{0.5cm}and \hspace{0.5cm} $U\equiv$ \includegraphics[width=1cm]{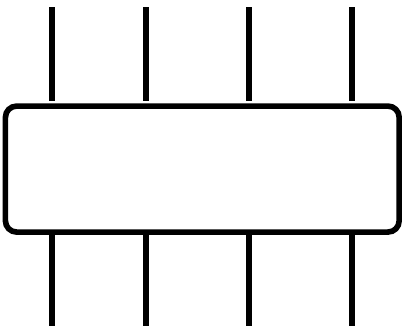} \hspace{0.5cm}
      \mbox{so that} \quad $U|\psi\rangle \equiv\, $\,\includegraphics[width=1cm]{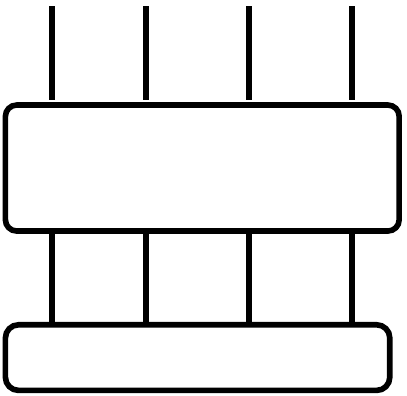} 
    \caption{Notation for tensor networks. 
    }
    \label{fig:fig1}
\end{figure}

The construction is most clearly described pictorially, using some standard notation for tensor networks. This notation, illustrated in Fig.~\ref{fig:fig1}, enables us to depict a state vector $|\psi\rangle$ and a component $\langle  \uparrow\uparrow\downarrow \downarrow |\psi\rangle$ of this state vector in a convenient (e.g. computational) basis, as objects with a single set of legs (the thin, vertical lines in these pictures), which may carry component labels referring to the local Hilbert space basis. Conversely, an operator, such as the unitary matrix $U$, carries two sets of legs, corresponding to the two indices of a rank-two tensor. As usual, the action $U|\psi\rangle$ of this matrix (a quantum gate) on a state vector is computed by contraction of indices, and this contraction is implied for all internal legs in these pictures. 

\begin{figure}[htb]
    \centering 
    \includegraphics[width=5cm]{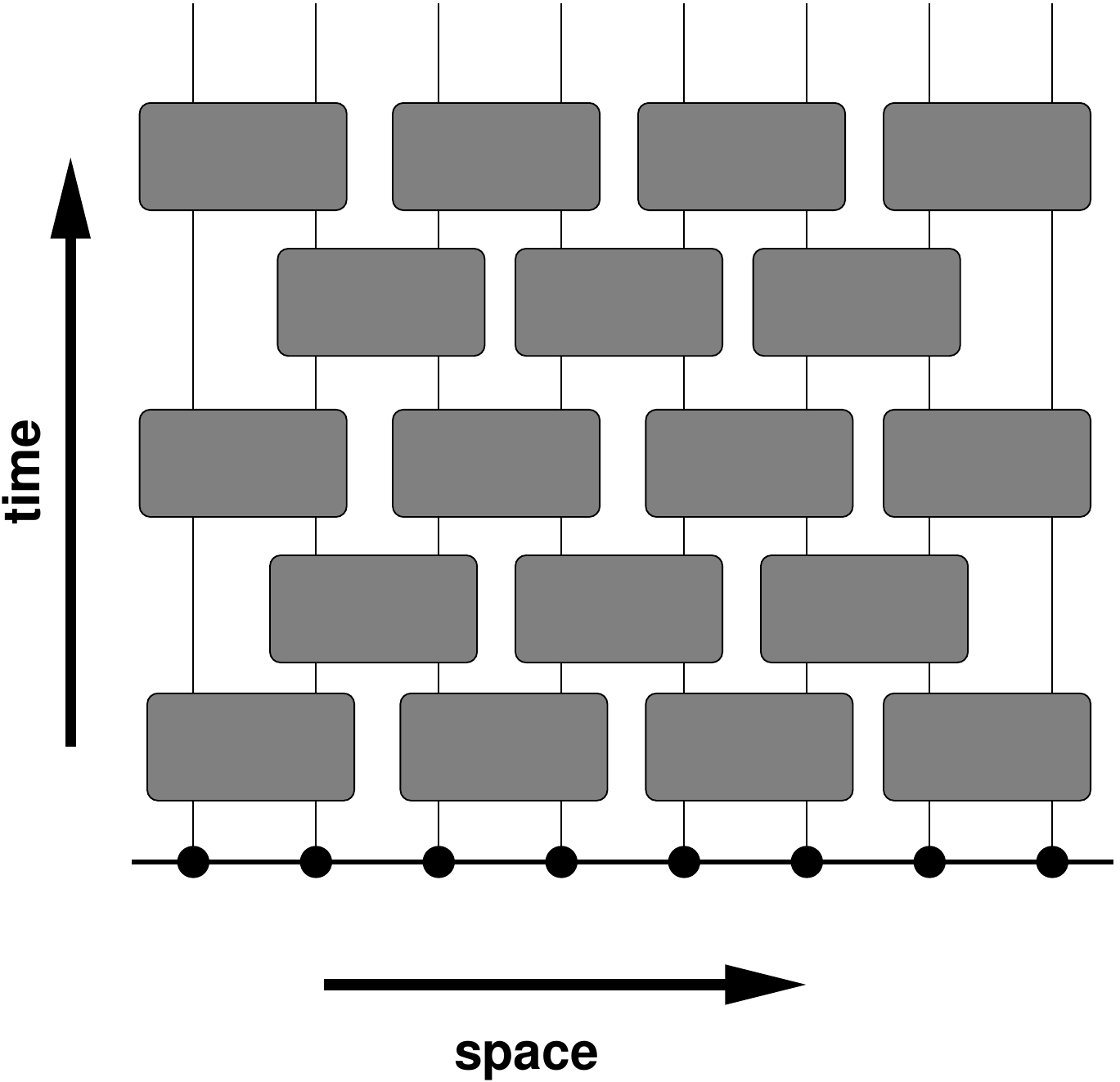}
    \caption{A brickwork quantum circuit. 
    }
    \label{fig:fig2}
\end{figure}

A local quantum circuit is built by combining the actions of many gates, each acting only on nearby sites \cite{Nielson_2000}. For simplicity, we will restrict ourselves to gates that act on pairs of sites, represented by $q^2\times q^2$ unitary matrices. One way to do this, using the so-called brickwork architecture, is illustrated in Fig.~\ref{fig:fig2}. This picture, drawn with space as the horizontal coordinate and time as the vertical one, shows time-evolution of an initial state (at the bottom of the figure), which proceeds in discrete steps. Over a single step, neighbouring pairs of sites are coupled by gates, which act between alternate sets of neighbours at successive steps, so that all parts of the system are eventually coupled after sufficiently long evolution. To obtain an ensemble of random unitary circuits (RUCs), each gate is chosen independently from the Haar distribution \cite{NahumQuantumentanglement}. The resulting construction is similar in appearance to the Trotter decomposition, used to approximate evolution by a time-independent Hamiltonian, but with the important difference that in the Trotter decomposition one is interested in the limit in which all gates are close to the identity, which is not the case for Haar gates.

More formally, consider the evolution operator $W(t)$ for a system of $L$ sites acting over the time interval $(0,t)$, which is a $q^L\times q^L$ unitary matrix. Taking $L$ to be even for convenience, we set $W(t) = W_tW_{t-1} \ldots W_1$ with each $W_\tau$ of the form $W_\tau = U_1\otimes U_2 \otimes \ldots \otimes U_{L/2}$ if $\tau$ is odd, and of the form $\mathbbm{1}\otimes U^\prime_1 \otimes U^\prime_2 \ldots U^\prime_{L/2-1}\otimes \mathbbm{1}$ if $\tau$ is even. Here, each $U_k$ and $U^\prime_k$ is a $q^2 \times q^2$ unitary matrix drawn from the Haar distribution, independently for each $k$ and each $\tau$. 

By this route, we have a model for many-body quantum dynamics that is minimally structured: it builds in locality but beyond this is arguably as simple as possible \cite{NahumQuantumentanglement}. In essence, the programme we outline in these notes is the calculation of physical properties of these or similar circuits, averaged over the ensemble.

Of course, the same time evolution can also be generated by the action of a Hamiltonian, which must be time-dependent to reflect the presence of discrete gates. The absence of continuous time-translation symmetry (and of other symmetries if the gates are Haar random) implies that the model has no conserved local densities (not even energy density, as would be the case for a time-independent, local Hamiltonian). This absence of conservation laws is itself an indication of the simplicity of RUCs.

There are many ways of extending this basic picture. Some of these involve re-introducing symmetries, with different outcomes according to the motivation. In one case, motivated by interest in the statistical properties of spectra for generic quantum systems, we would like to have a fixed evolution operator, defining a definite spectrum, rather than an evolution operator that changes randomly in time, as happens for the RUC defined above. We can do this by considering an ensemble of random Floquet circuits (RFCs), in which the gate acting between a given neighbouring pair of sites is taken to be the same Haar random unitary each time, while statistically independent of the gates acting between other neighbouring pairs. Formally, we write $W(t) = W^t$, where $W$ is the Floquet operator, giving evolution over a single period. For a brickwork RFC, we take $W=W_2\cdot W_1$ with $W_1$ and $W_2$ defined as for an RUC. In this way we impose discrete time-translation invariance. Since at late times, each two-site unitary gate appears many times in physical quantities, calculations of ensemble-averaged properties are more difficult for RFCs than for RUCs, and require the simplification of large $q$ \cite{Chan_PRX}. 

As a second case, we may want to retain the Markov nature of an RUC but introduce a continuous symmetry, and with it a conserved density. Taking this symmetry to be U(1), one can enlarge the local Hilbert space dimension from $q$ to $2q$, replacing $\mathbb{C}^q$ with $(\uparrow,\downarrow)\otimes \mathbb{C}^q$ \cite{Khemani_2018,Rakovsky_2018}. Then the total $z$-component of spin is conserved by taking two-site gates to be block-diagonal $(2q)^2\times (2q)^2$ unitaries of the form
$$
\begin{array}{cccc} 
\qquad \uparrow\uparrow&\quad
\uparrow\downarrow&\quad
\downarrow\uparrow&\quad
\downarrow\downarrow
\end{array}
$$
$$
\begin{array}{c}\uparrow\uparrow\\
\\
\uparrow\downarrow\\
\\
\downarrow\uparrow\\
\\
\downarrow\downarrow
\end{array}
\left(\begin{array}{ccc}
U_{q^2\times q^2} & &\\ 
& & \\ &&\\
& U_{(2q^2)\times (2q^2)} & \\ 
& & \\ && \\
& & U_{q^2\times q^2}
\end{array}\right)
$$
with each of the three blocks (two of size $q^2\times q^2$ and one of size $2q^2\times 2q^2$) independent Haar random unitary matrices.

\begin{figure}[htb]\label{fig:fig3}
    \centering 
    \includegraphics[width=8cm]{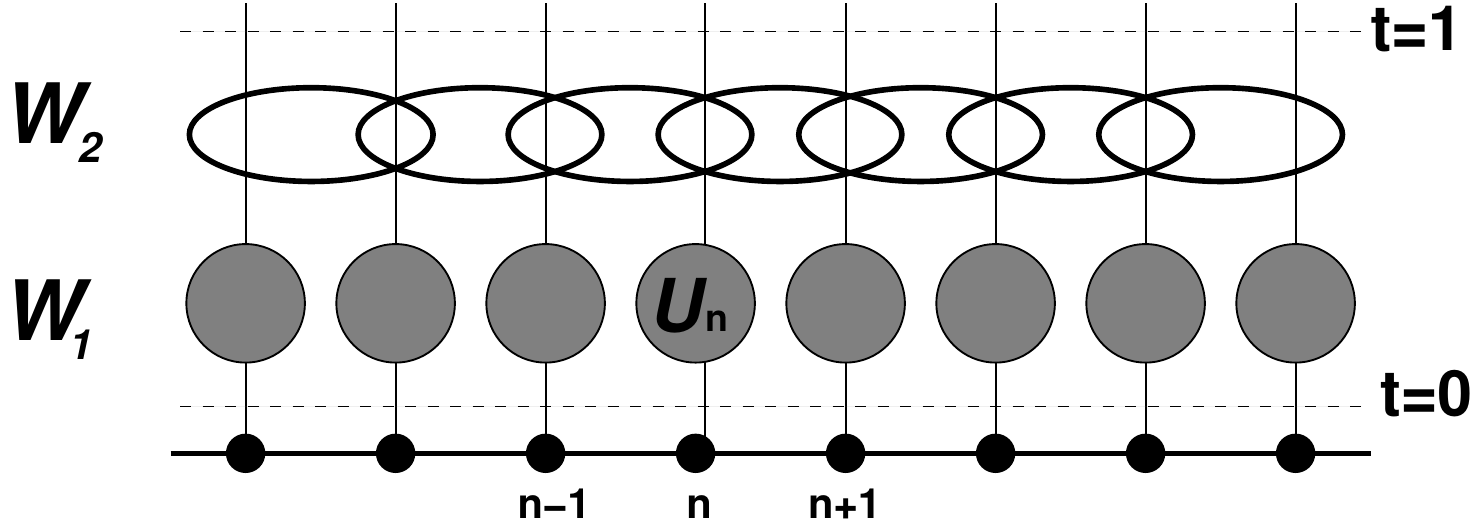}
    \caption{The random phase model 
    }
\end{figure}

Some other useful extensions do not follow simply from symmetry considerations. One of these is to introduce a parameter that controls the strength of coupling between sites, using the random phase model (RPM) \cite{Chan_PRL}. This is valuable for RFCs because in the case of a brickwork circuit with Haar unitaries, the large $q$ limit necessary for analytic calculations results in a system with strong coupling between sites, which obscures some of the physical phenomena that we observe in numerics at finite $q$. One Floquet timestep of the RPM is illustrated in Fig.~\ref{fig:fig3}.
In equations, the Floquet operator is defined as $W=W_2\cdot W_1$ with $W_1 = U_1\otimes U_2 \otimes \ldots U_L$, where each $U_k$ is a $q\times q$ independent Haar-random unitary matrix that makes rotations within the local Hilbert space on site $k$. The factor $W_2$ is diagonal in the computational basis $|a_1,a_2 \ldots a_L\rangle$ with phases $\sum_{n=1}^{L-1} \varphi(a_n,a_{n+1})$, which are sums of independent Gaussian random variables with mean zero and variance $[\varphi(a_m,a_{m+1})\varphi(b_n,b_{n+1})]_{\rm av} = \varepsilon\,\delta_{mn}\delta_{a_m,b_n} \delta_{a_{m+1},b_{n+1}}$. The value of $\varepsilon$ controls the strength of coupling between sites: they are uncoupled if $\varepsilon=0$ but strongly coupled if $\varepsilon \gtrsim 1$. As in any Floquet model, the time evolution operator $W(t)$ over $t$ timesteps is given by raising the Floquet operator $W$ to the $t$-th power.

Other variations are also useful in some settings. For example, one can introduce randomness in the space-time architecture of the circuit, by choosing at random for each time-step which neighbouring pair of sites to couple with the next gate added to the circuit. Such additional randomness, beyond the realisation of the unitary matrix corresponding to each gate, may help to illustrate features of entanglement spreading that hold more generally \cite{NahumQuantumentanglement}.

\section{Operator spreading}

\begin{figure}[htb]
    \centering 
    \begin{minipage}{0.3\textwidth}
    \includegraphics[width=1.5cm]{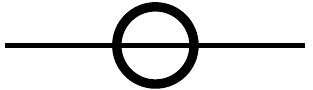}\\
    \vspace{0.5cm}
    ~\\
    \includegraphics[width=1.5cm]{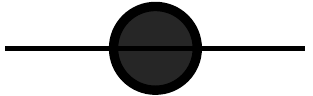}
    \end{minipage}
    \begin{minipage}{0.6\textwidth}
 \includegraphics[width=5cm]{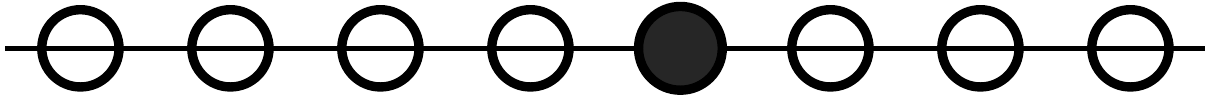}\\
  \includegraphics[width=5cm]{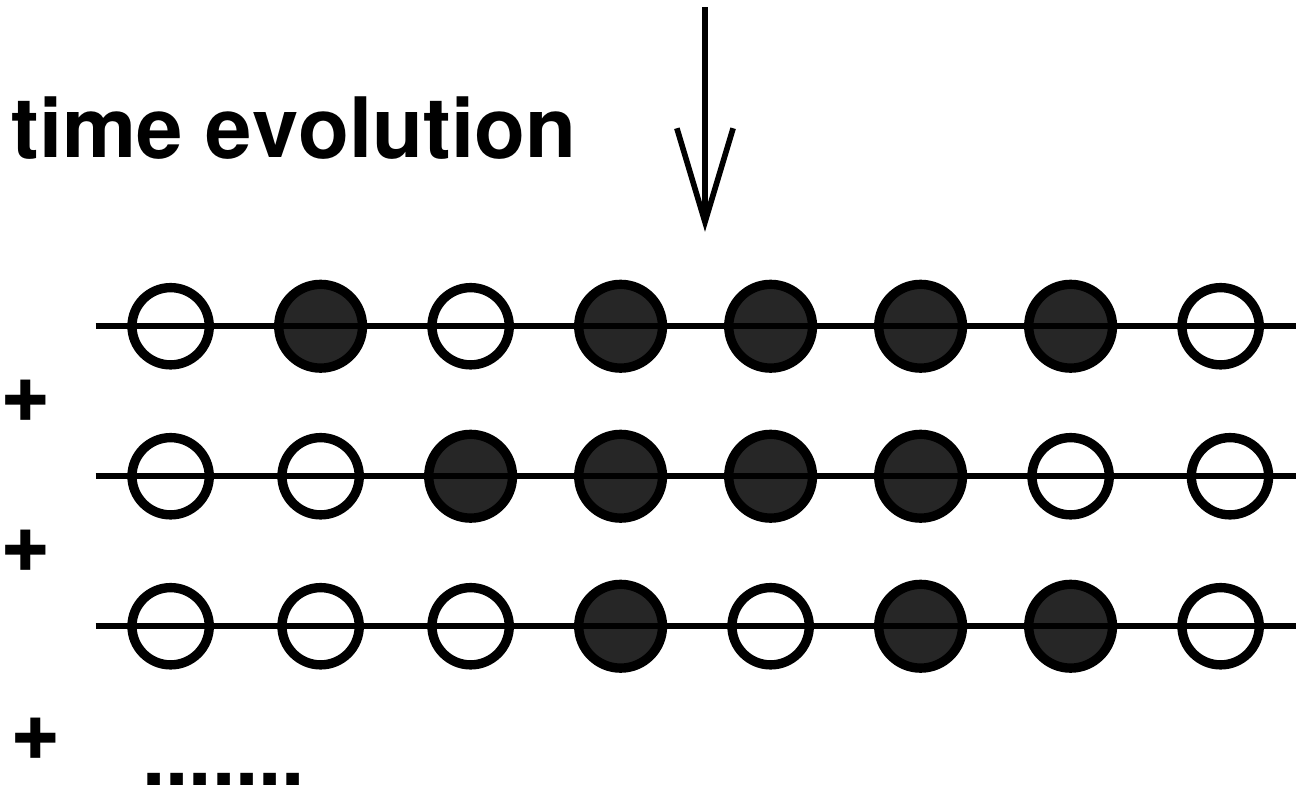}
\end{minipage}
    \caption{An illustration of operator spreading. (Left) Notation: an open circle represents a site at which the operator acts trivially [$\mathbbm{1}$ in Eq.~\eqref{eq:opspread}] and a filled circle represents a site at which the operator acts non-trivially [$X$, $Y$ or $Z$ in Eq.~\eqref{eq:opspread}]. (Right) Time-evolution: (top) $O(x)$ and (bottom) $O(x,t)$.
    }
    \label{fig:fig4}
\end{figure}

As a first exercise in using RUCs, we will discuss spreading of an initially local operator under time evolution in the Heisenberg picture, with a presentation that follows closely the original papers \cite{vonKeyserlingk_2018,NahumSpreading}. Let $O(x)$ be a Hermitian operator that acts locally on the system at the site $x$. We want to discuss the time-evolved operator
\begin{equation}
    O(x,t) = W^\dagger(t) O(x) W(t)
\end{equation}
with $W(t)$ the time-evolution operator for a circuit, as introduced above. To be specific we can specialise to the case $q=2$ and use the notation $X$, $Y$ and $Z$ for the three Pauli matrices. Then with a specific choice for $O(x)$ and writing this explicitly as an operator acting in the $q^L$-dimensional Hilbert space of the full system, rather than just the local Hilbert space at site $x$, we might have
\begin{equation}
    O(x) = \mathbbm{1} \otimes \mathbbm{1} \otimes X \otimes \mathbbm{1} \ldots \,.
\end{equation}
The idea of operator spreading is the anticipation that $O(x,t)$ may include a contribution of the form (say)
\begin{equation}\label{eq:opspread}
    \mathbbm{1}\otimes Y \otimes Z \otimes Y \otimes \mathbbm{1} \ldots
\end{equation}
as well as many others. This is illustrated schematically in Fig.~\ref{fig:fig4}

In order to develop a more formal treatment, we must introduce some definitions. We will view operators as elements of a vector space. To do so we must define a scalar product, and for Hermitian operators $\mathcal{A}$ and $\mathcal{B}$ acting on an $N$-dimensional Hilbert space we take this to be $N^{-1}{\rm Tr}[\mathcal{A}\mathcal{B}]$. In general, there are $N^2$ linearly independent such operators. In our case, we can define a complete orthonormal basis for the full $q^L$-dimensional vector space of the system by first defining a complete orthonormal operator basis at each site, and then taking direct products of these elements. Since we want to be able to distinguish sites on which operators act non-trivially, we should include the identity as one element of the basis at each site. These basis elements for the system as a whole are known as operator strings and we denote them with $\mathcal{S}$ (using this symbol both for the string itself and as a label). We can then write
\begin{equation}
    O(x,t) = \sum_{\mathcal S}a_{\mathcal{S}}(t)\mathcal{S}
\end{equation}
where the amplitudes $a_{\mathcal{S}}(t)$ can be shown to be real numbers, provided the operators $\cal S$ and $O(x)$ are Hermitian. This expansion for a time-evolving operator has close parallels with the usual expansion of a time-evolving wavefunction in terms of a set of basis states. In particular, there is a conserved probability density associated with the amplitudes, since (noting that ${\rm Tr}[O(x,t)^2] ={\rm Tr}[O(x)^2]$, choosing a normalisation for the initial operator by setting $q^{-L}{\rm Tr}[O(x)^2]=1$, and using $q^{-L}{\rm Tr}[\mathcal{S}\mathcal{S}^\prime] = \delta_{\mathcal{S},\mathcal{S}^\prime}$) we have
\begin{equation}
   \sum_{\mathcal{S}} [a_{\mathcal{S}}(t)]^2 =1
\end{equation}
for all $t$. 

Knowledge of the probabilities $[a_{\mathcal{S}}(t)]^2$ for all $\mathcal{S}$ would provide a great deal of information about the system, but we content ourselves with a reduced view by organising the strings according to the location of (say) their right-most non-identity operator. (Of course, we could alternatively focus on the left-most end, with equivalent results). To make this reduction, we sum the probabilities of all strings that have a non-identity operator at site $k$, but only identity operators to the right of this, defining
\begin{equation}
    p_k(t) = \sum^\prime_{\mathcal{S}}\,[a_{\mathcal S}(t)]^2 \,.
\end{equation}
Clearly, the $p_k(t)$ in turn define a normalised probability distribution, since $\sum_k p_k(t) =1$ for all $t$.

\begin{figure}[htb]
    \centering 
    \includegraphics[width=12cm]{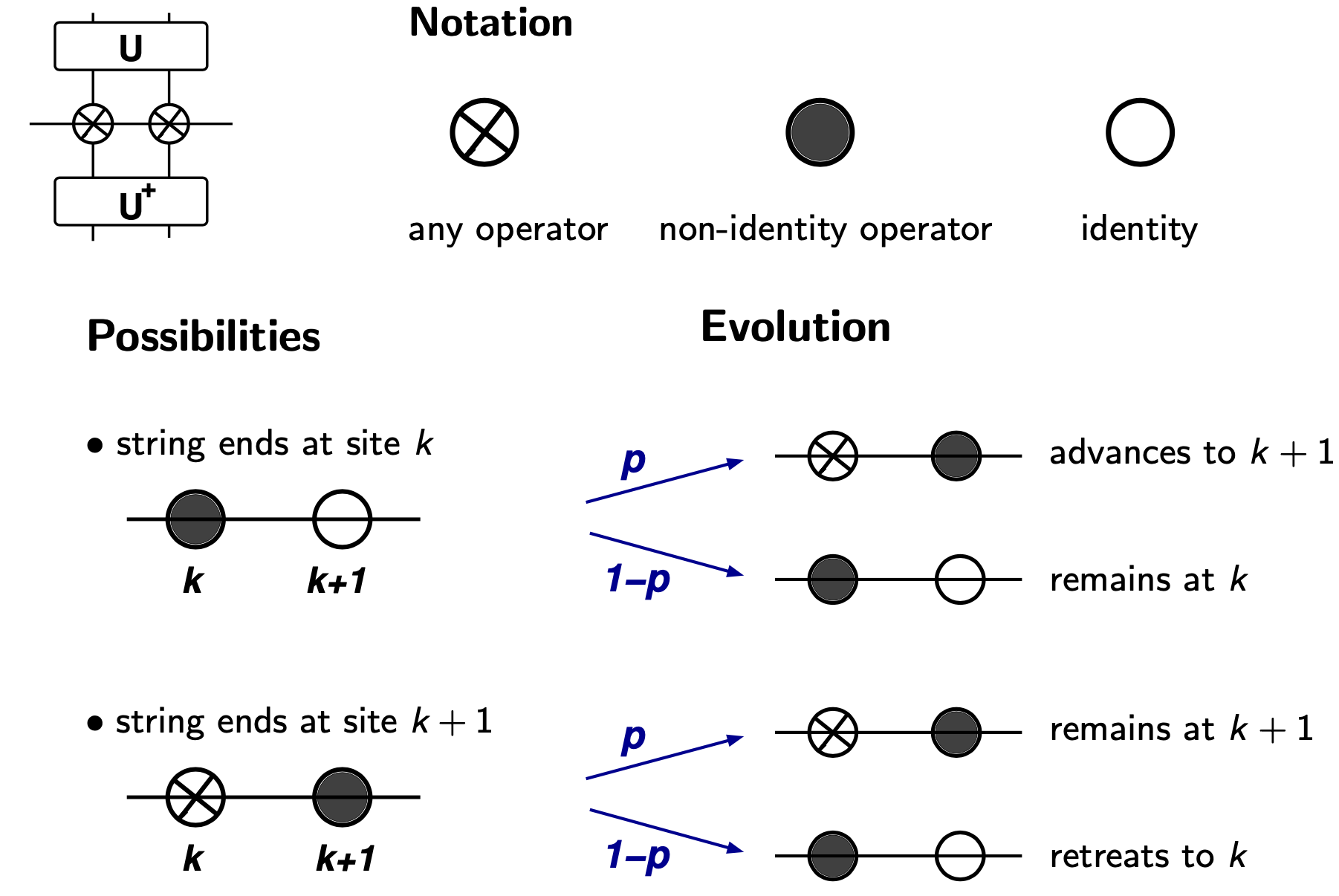}
    \caption{Schematic illustration of operator evolution under the action of a two-site gate.
    }
    \label{fig:fig5}
\end{figure}

Our aim now is to study the ensemble-averaged evolution of $p_k(t)$ with $t$ in an RUC, and we will see that this is represented by a Markov process of a type familiar in classical statistical physics. Since each step of the time evolution for the circuit involves the action of two-site gates, it is sufficient to consider evolution just of the part of an operator string involving only two sites, with the string ending at one or other of these two sites. This evolution is represented pictorially in Fig.~\ref{fig:fig5}, in which we extend slightly the notation used in Fig.~\ref{fig:fig4}. Considering the action of a gate that couples sites $k$ and $k+1$, there are two initial operator configurations that are of interest to us because from these configurations the action of this gate may change the location of the end of the operator string. In one, the string ends at site $k$, in which case the operator on $k$ is not the identity while the operator on site $k+1$ is necessarily the identity. In the other, the string ends at site $k+1$, so that the operator there is not the identity, while the operator at site $k$ may be anything. It is evident that the action of the two-site gate may take these initial configurations to the final ones listed on the right-hand side of Fig.~\ref{fig:fig5}, where we have used $p$ and $(1-p)$ to denote the weights for each outcome, after averaging over the ensemble of two-site gates (at this point, it is not obvious that the same $p$ arises for both initial configurations). 

The next step is to verify the picture given in Fig.~\ref{fig:fig5} and compute the value of $p$. This can be done explicitly by taking the scalar product of one of the possible final operator configurations in Fig.~\ref{fig:fig5} with the time-evolved operator, squaring the result, then averaging on $U$ using results for Haar averages that will be outlined in Sec.~\ref{sec:entanglement}. 

\begin{figure}[htb]
    \centering 
    \includegraphics[width=5cm]{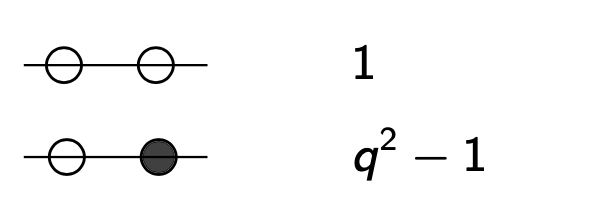}\hspace{1cm}
    \includegraphics[width=5cm]{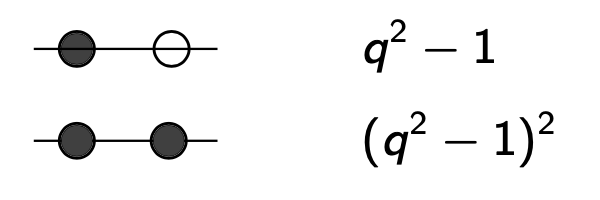}
    \caption{Illustration of numbers of two-site operators of different types.
    }
    \label{fig:fig6}
\end{figure}

We will avoid this calculation by instead using simple arguments to guess its outcome. The central idea is that, given the Haar average, all possible outcomes are equally likely, and so we simply need to count them. For a single site, there are $q^2$ operators in the basis, of which one is the identity $\mathbbm{1}$ and $q^2-1$ are orthogonal to the identity. Similarly, for two sites there are $q^4$ operators in total, of which $q^4-1$ are orthogonal to the two-site identity $\mathbbm{1}\otimes\mathbbm{1}$. Now a crucial point is that evolution from either of the starting operators shown in Fig.~\ref{fig:fig5} cannot yield the two-site identity. To see this, consider following the operator evolution backwards in time, by computing $U (\mathbbm{1}\otimes \mathbbm{1})U^\dagger = \mathbbm{1}\otimes \mathbbm{1}$. In other words, the identity can only be reached as the final operator by evolution from the identity, which is not one of the initial operators included in Fig.~\ref{fig:fig5}. We therefore have $q^4-1$ possible linearly independent final operators on the right-hand side of Fig.~\ref{fig:fig5}, which can be subdivided as illustrated in Fig.~\ref{fig:fig6}. 

Consider as an example the evolution that is depicted in the upper line of Fig.~\ref{fig:fig5}, which takes an initial string ending at site $k$ to a final string ending at site $k+1$. We see that this can happen in $q^2(q^2-1)$ ways out of $q^4-1$ possibilities in total. Hence
\begin{equation}
    p = \frac{q^2(q^2-1)}{q^4-1} = \frac{q^2}{q^2+1}\,.
\end{equation}
Other cases can be treated in a similar way, confirming the same value of $p$ at each appearance in Fig.~\ref{fig:fig5}. In summary, under RUC dynamics, and after averaging on Haar-distributed gates, the right-hand string end undergoes a random walk, which crucially is biased to the right since $1/2<p\leq 1$ for all physical $q$.

\begin{figure}[htb]
    \centering 
    \includegraphics[width=8cm]{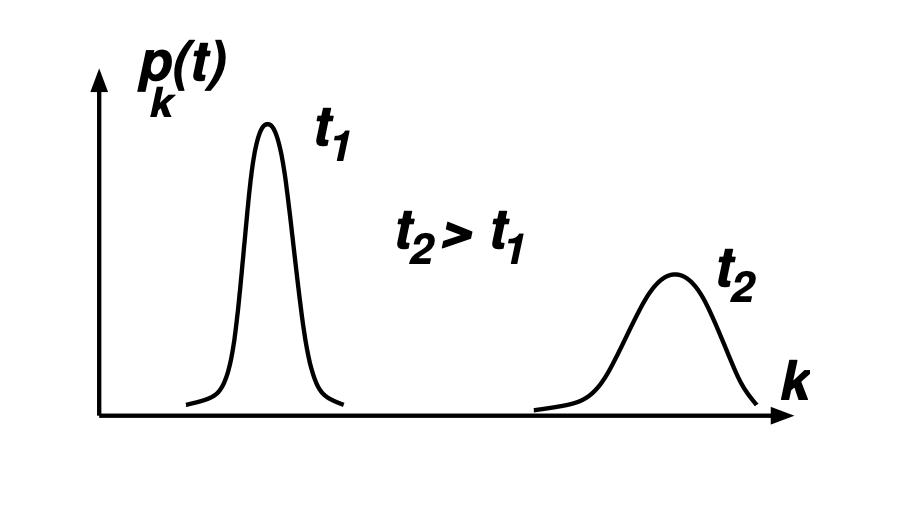}
    \caption{Evolution with time $t$ of the probability $p_k(t)$ for the right-hand end of an operator string to lie at site $k$.
    }
    \label{fig:fig7}
\end{figure}
It is straightforward to write down and solve explicitly a master equation for the dynamics \cite{vonKeyserlingk_2018,NahumSpreading} but we will limit ourselves to a discussion of the main results. The probability distribution $p_k(t)$ evolves as sketched in Fig.~\ref{fig:fig7}. In particular, at large $t$ the mean location of the string end for an operator initially acting at the origin satisfies
\begin{equation}
    \langle k(t) \rangle \equiv \sum_k k\, p_k(t) \simeq \langle k(0) \rangle + vt \quad \mbox{with} \quad v = 2p-1 = \frac{q^2 - 1}{q^2+1}
\end{equation}
while its position around this point broadens diffusively, with
\begin{equation}
    \langle k(t)^2\rangle - \langle k(t) \rangle^2 \simeq 2Dt \quad \mbox{with} \quad D = \frac{2q^2}{(q^2+1)^2}\,.
\end{equation}
In summary, an operator that is initially simple, in the sense of acting locally, becomes on average more complicated under time evolution, in the sense that it evolves into a superposition of operator strings that have a length growing linearly in time. This growth is a description of the process of equilibration within the Heisenberg picture.  For example, the autocorrelation function 
\begin{equation}
    \langle O(x,t)O(x) \rangle \equiv q^{-L}{\rm Tr}[O(x,t)O(x)]
\end{equation}
is the amplitude of the initial operator in the evolving superposition. This amplitude decays with increasing $t$ because almost all weight is transferred to longer strings. See Refs.~\cite{vonKeyserlingk_2018,NahumSpreading,Yoshimura_2025} for further details.

While information about $a_{\cal S}(t)$ and $p_k(t)$ gives a rather detailed picture of operator spreading, it is also useful to have a quantity defined in the spirit of a correlation function, that summarises some of the same information. One such quantity is known as the out-of-time-order correlator (OTOC), $N^{-L}{\rm Tr} \, |[O(x),O(y,t)]|^2$. Here $O(x)$ and $O(y)$ are operators acting at points $x$ and $y$ in the system, $O(y,t)$ is time-evolved in the Heisenberg picture, and we consider the thermal average of the squared magnitude of their commutator, using the infinite temperature density matrix since that is appropriate in a driven system such as an RUC. The magnitude of the commutator probes operator spreading: if the points $x$ and $y$ are well separated, then we expect at early times that the operators $O(x)$ and $O(y,t)$ will be supported on disjoint parts of the system and will therefore commute; conversely, at long times, operator spreading leads to an overlap in their supports and so the commutator should be non-zero. 

Rigorous and rather general bounds on operator spreading from this perspective were established in early work by Lieb and Robinson \cite{Lieb_1972}. In our notation and setting, the bound has the form
\begin{equation}\label{LR}
|| [O(x),O(y,t)]|| \leq C e^{-a(|x-y| -v|t|)}
\end{equation}
where $C$, $a$ and $v$ are real, positive model-dependent constants. The significance of Eq.~\eqref{LR} is that, if we take $|x-y|\to\infty$ and $|t|\to\infty$ along a ray with $\tilde{v} \equiv |x-y|/|t|$ fixed, then the commutator tends to zero provided $\tilde{v} > v$. The OTOC is intended to access this behaviour via the square of the commutator, which is a much simpler quantity to handle that the operator norm appearing in Eq.~\eqref{LR}.

It is useful to make some simplifying choices and to isolate the part of the OTOC of most interest, as follows. We take $O(x)$ and $O(y)$ to be traceless operators that square to give the identity (as is the case for the Pauli spin operators), and we expand the commutator, writing
\begin{eqnarray}
    \frac{1}{2}N^{-L}{\rm Tr} \, |[O(x),O(y,t)]|^2 &=& N^{-L}{\rm Tr}\left\{[O(x)^2 O(y,t)^2] \right. \nonumber- \left.[O(y,t) O(x) O(y,t) O(x)] \right\}\nonumber\\
    &=& 1 - C(x,y;t) \nonumber\\
    \mbox{with} \quad C(x,y;t) &\equiv& N^{-L}{\rm Tr}\,[O(y,t) O(x) O(y,t) O(x)]\,.
\end{eqnarray}
The important difference between $C(x,y;t)$ and the correlation functions that characterise an equilibrium system within, for example, linear response theory is that in the latter case operators appear in time order, whereas in $C(x,y;t)$ they do not. At early times, as discussed, the operators $O(x)$ and $O(y,t)$ commute, and so we have the simplification 
$$
C(x,y,t) \approx N^{-L} {\rm Tr} [O(y,t)^2 O(x)^2] = 1\quad \mbox{for} \,\,t\,\,\, \mbox{small}.
$$
At late times, when almost all the operator strings that make up $O(y,t)$ extend past the point $x$, we expect 
$$
C(x,y,t)\to 0 \quad \mbox{for} \,\,t\,\,\mbox{large}
$$
from the following argument. First, consider changing the order of operators in $C(x,y;t)$, from $O(y,t) O(x) O(y,t) O(x)$ to $O(y,t)^2 O(x)^2$. Taking $q=2$ for simplicity, so that the operators act as Pauli matrices or the identity at each site, each string either commutes or anticommutes with $O(x)$. Haar evolution ensures that both possibilities contribute with equal weight, yielding zero on average. 

\begin{figure}[htb]
    \centering 
    \includegraphics[width=5cm]{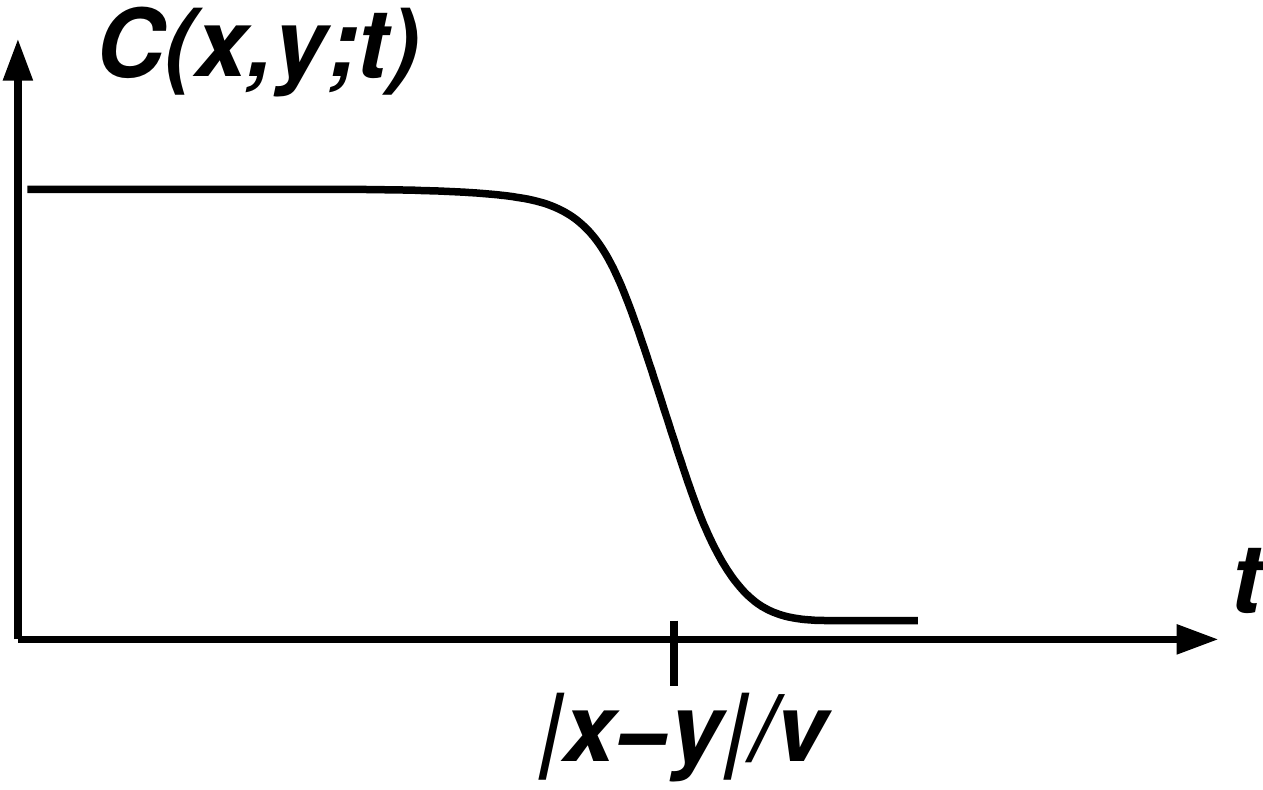}\hspace{2cm}
    \includegraphics[width=5cm]{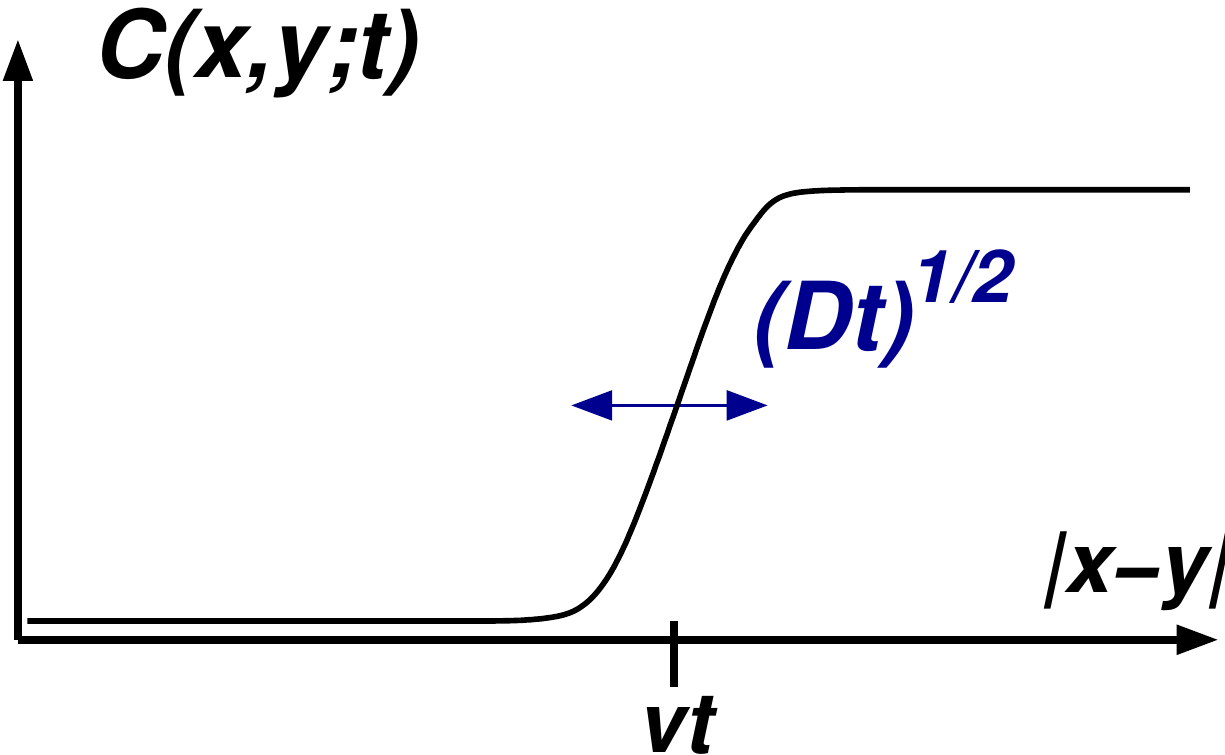}
    \caption{Behaviour of the correlator $C(x,y;t)$: (left) as a function of time $t$ at fixed separation $|x-y|$; (right) as a function of separation at fixed time.
    }
    \label{fig:extra}
\end{figure}

There is an explicit connection between the correlator $C(x,y;t)$ and the probability distribution $p_k(t)$ is string ends. To derive it, we separate contributions to $O(y,t)$ from strings of different length, distinguishing between ones that extend beyond the point $x$ and ones that do not, since only the former contribute to the commutator. Moreover, for these contributing operator strings, all possible operators appear with equal weight at site $x$. As a result, one finds \cite{vonKeyserlingk_2018,NahumSpreading}
\begin{equation}
    C(x,y;t) = \sum_{k<x} p_k(t)
\end{equation}
which leads to the behaviour for $C(x,y;t)$ shown in Fig.~\ref{fig:extra}.

\section{Entanglement dynamics}\label{sec:entanglement}

As a second illustration of the understanding that comes from calculations on RUCs, we will discuss spreading of spatial entanglement under time evolution from an initial state that is direct product in the computational basis. Again, our presentation follows closely the original papers \cite{NahumQuantumentanglement,NahumEntanglementmembrane}. 

\begin{figure}[htb]
    \centering 
    \includegraphics[width=8cm]{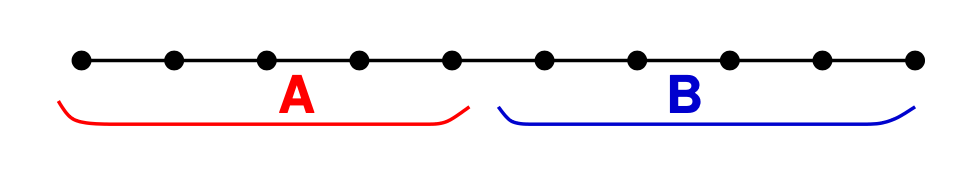}
    \caption{Division of the system into subsystems, as used in the definition of the reduced density matrix.
    }
    \label{fig:fig8}
\end{figure}

We start by recalling some elementary features of the density matrix description of a quantum system. For a system in a pure quantum state $|\psi(t)\rangle$ the density matrix is $\rho(t) = |\psi(t)\rangle \langle \psi(t)|$. One of its eigenvalues is unity and the remainder are zero, and this feature is preserved under time evolution. To see signatures of equilibration at long times, it is necessary to consider a reduced density matrix for a sub-system. To this end, suppose the full system is divided into two parts, labeled $A$ and $B$ in Fig.~\ref{fig:fig8}  (with lengths $L_A$ and $L_B$, and $L_A+L_B=L$), and let us decide only to consider observations on sub-system $A$. Their outcomes can be computed from the reduced density matrix for that subsystem, $\rho_A(t) = {\rm Tr}_B [\rho(t)]$. Crucially, although $\rho(t)$ describes a pure state at all times, the reduced density matrix is in general mixed. Its eigenvalues (which are real) can be ordered as $\lambda_1 \geq \lambda_2 \geq \ldots \lambda_{q^{L_A}}\geq 0$ with $\sum_k \lambda_k = 1$ and we would like to understand how they evolve with time. For example, if the initial state is a product state in a site basis, then $\rho_A({t=0})$ is pure, so that initially $\lambda_1 = 1$ and $\lambda_k =0$ for $k>1$. Conversely, in equilibrium at infinite temperature (the maximally mixed state)  we have $\lambda_k = q^{-L_A}$ for all $k$. We would like to investigate whether time evolution takes the system from the first behaviour to the second one.

Information about the distribution of the eigenvalues of the reduced density matrix requires study of non-linear functions of $\rho_A(t)$ since the value of the linear function $\sum_k\lambda_k$ is fixed. The simplest such quantity is the purity ${\rm Tr}[\rho_A(t)]^2$. More generally, one can consider the R\'enyi entropies $S_\alpha$ defined by
\begin{equation}
    e^{-(\alpha - 1) S_\alpha(t)}={\rm Tr}[\rho_A(t)^\alpha]\,.
\end{equation}
The purity is then $e^{-S_2(t)}$ while the von Neumann entropy is given by 
$$
\lim_{\alpha \to 1}S_\alpha(t) = - {\rm Tr }[\rho_A(t) \ln \rho_A(t)].
$$ 
Note that in the limiting case of a pure state, $S_\alpha =0$ for all positive $\alpha$, and in the opposite limit of the maximally mixed state, $S_\alpha = L_A \ln q$. 

We will discuss two approaches to understanding the time-dependence of the R\'enyi entropies for a system initially in a product state in a site basis. In both cases we follow Ref.~\cite{NahumQuantumentanglement}. In the first approach (see also, for example, Refs.~\cite{Casini_2016} and \cite{Swingle_2012}), the idea is to get an upper bound on the rank of the reduced density matrix and examine how this changes with time. In the second approach, we set out an explicit calculation of the ensemble-averaged value of the purity of the reduced density matrix. 

\begin{figure}[htb]
    \centering 
    \includegraphics[width=8cm]{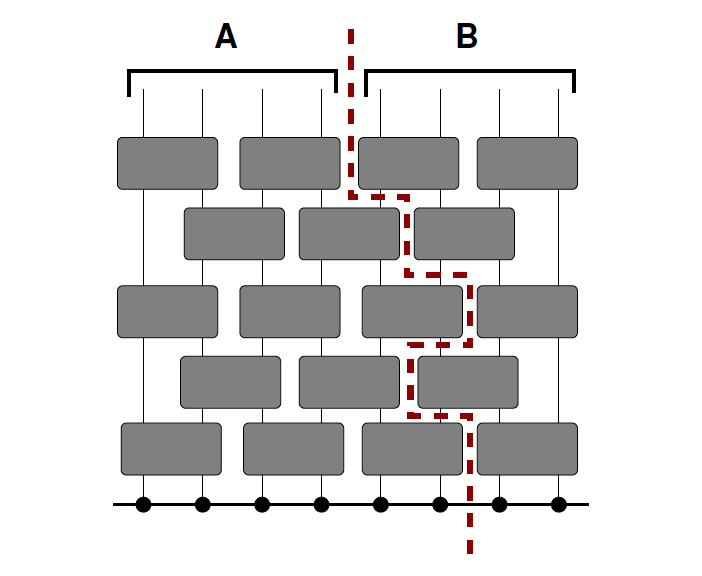}
    \caption{Illustration of the construction used to bound the rank of the density matrix in the \emph{min-cut} approach.
    }
    \label{fig:min-cut}
\end{figure}

To examine the rank of the reduced density matrix, we refer to Fig.~\ref{fig:min-cut}. Here we take a quantum circuit representing evolution for time $t$ from an initial product state, and draw a path, represented by the red dashed line in the figure, which runs from the spatial point dividing subsystems $A$ and $B$ at the final time, to exit the system either at the initial time or through the side of the system at an intermediate time. This path is drawn so that it runs between gates, but it necessarily cuts links in the circuit that connect gates. 

We now consider the outcome of summing over all state labels on links to the left of this path, while holding the labels fixed on the links crossed by the path. The result defines a state vector for subsystem $A$. Different choices of the labels on the links crossed by the path will yield other state vectors on the subsystem, and the reduced density matrix is given as a sum over outer products of all these vectors, each with a weight that depends on the part of the circuit to the right of the path. Although we have very little information about these state vectors, we know that their number sets an upper bound on the rank of the reduced density matrix. With $\ell_{\rm {cut}}$ denoting the number of legs in the circuit cut by the path, this bound on the rank is $q^{\ell_{\rm {cut}}}$. Equivalently, we can say that we perform a singular value decomposition of the state vector for the system at the final time using this path, finding that $q^{\ell_{\rm {cut}}}$ an upper bound to the rank of the decomposition. In turn this sets a bound on the R\'eyni entropies, via $S_\alpha \leq \ln[\mbox{rank of}\, \rho_A]$.

\begin{figure}[htb]
    \centering 
    \includegraphics[width=4cm]{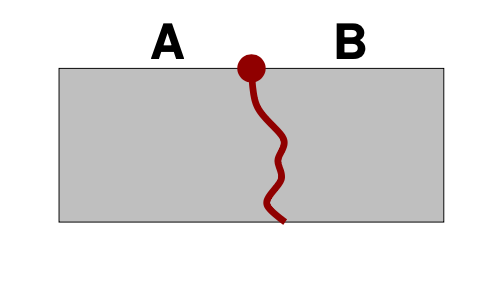}\hspace{2cm}
    \includegraphics[width=4cm]{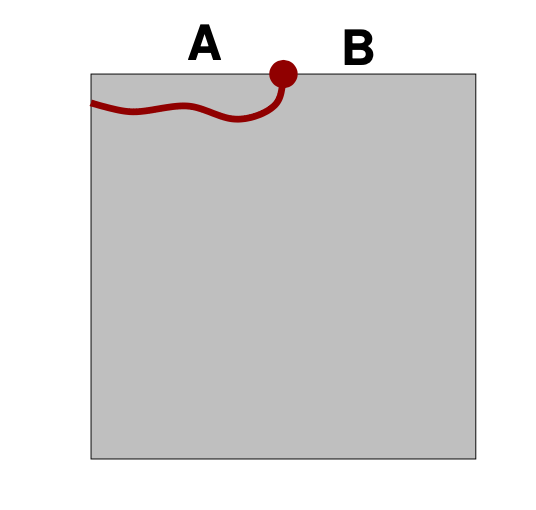}
    \caption{Dependence on time of the nature of the shortest path in the \emph{min-cut} approach. Left: at early times, the shortest path extends vertically to the initial state. Right: at sufficiently late times the shortest path exits the system laterally via the spatial boundary. 
    }
    \label{fig:fig13b}
\end{figure}

\begin{figure}[htb]
    \centering 
    \includegraphics[width=5cm]{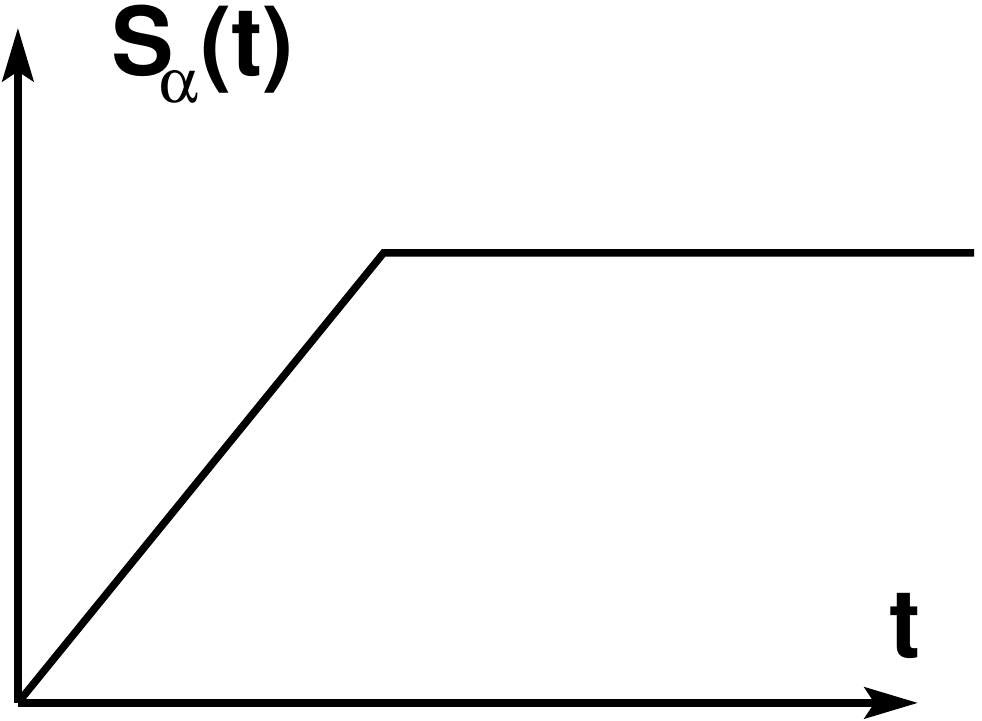}
    \caption{Dependence on $t$ of the upper bound for $S_\alpha$ derived from the minimum cut argument.
    }
    \label{fig:fig13a}
\end{figure}

To make the most of these bounds, we should choose the shortest path, with a length $\ell_{\rm min-cut}$. The nature of the shortest path depends on the geometry of the circuit, as shown in Fig.~\ref{fig:fig13b}. If the point dividing the subsystems lies far from the end of the system, then at early times the shortest path runs downwards, and $\ell_{\rm min-cut}$ grows (at least roughly) linearly in time. At late times, however, the shortest path leaves the system sideways at an intermediate time, and in this regime its length is independent of time. In consequence, this upper bound on $S_\alpha(t)$ has the dependence on $t$ sketched in Fig.~\ref{fig:fig13a}: it grows initially as a function of time, reaching a time-independent value at late times, as expected for an equilibration process.

\begin{figure}[htb]
    \centering 
    \includegraphics[width=13cm]{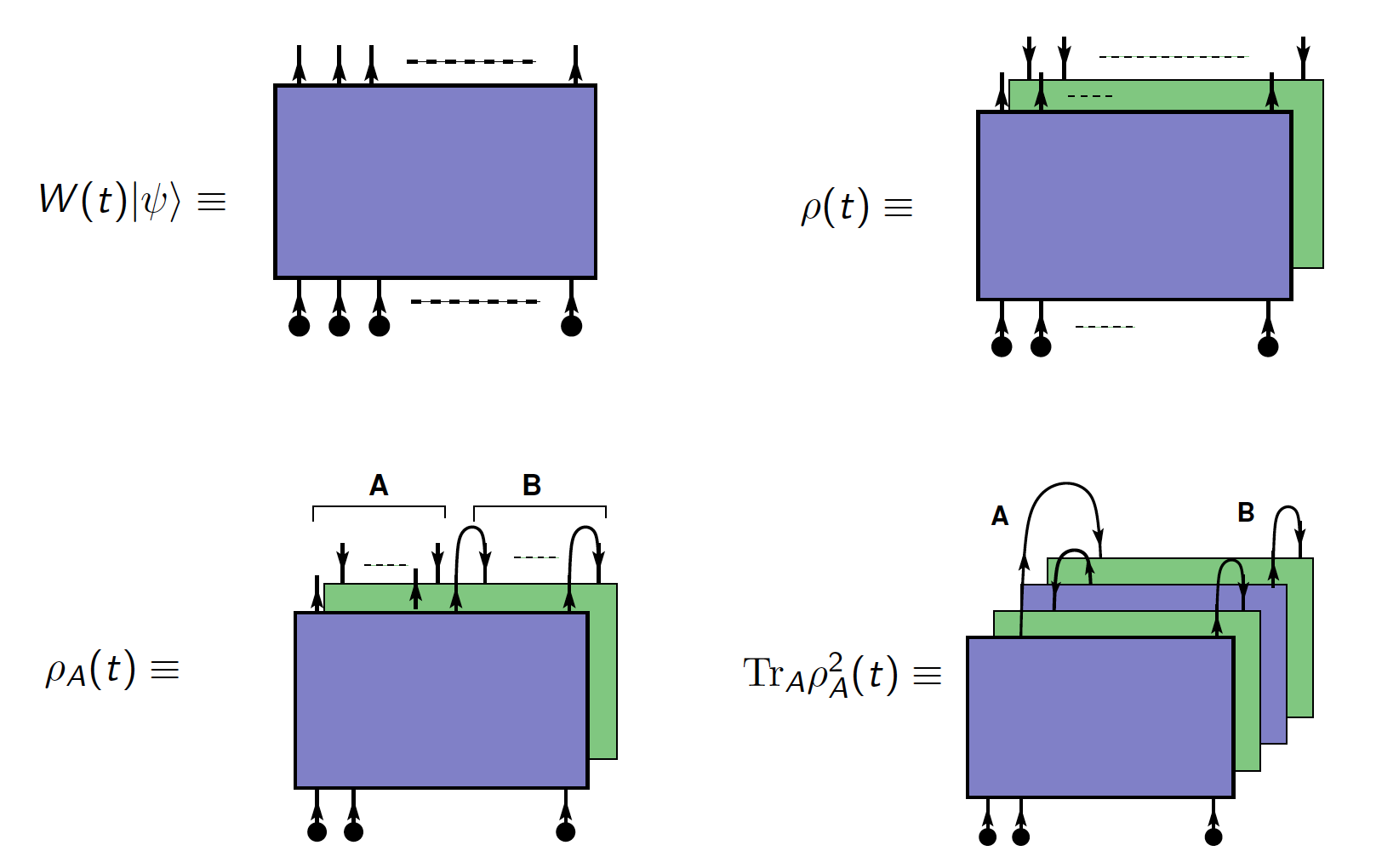}
    \caption{Schematic illustration of the steps required in a calculation of the purity ${\rm Tr}[\rho_A(t)]^2$. From the top left to the bottom right: (i) $|\psi(t)\rangle = W(t)|\psi\rangle$ is constructed by acting with $W(t)$ (blue rectangle) on an initial product state in the site basis (black dots at bottom of this panel); (ii) $\rho(t) = |\psi(t)\rangle \langle \psi(t)|$ (with $\langle \psi(t)|$ represented by a green rectangle); (iii) the reduced density matrix (where closed black lines indicate a trace); and (iv) the purity. 
    }
    \label{fig:fig9}
\end{figure}

Turning to a calculation of the purity, we can represent the steps required in a series of pictures, as shown in Fig.~\ref{fig:fig9} and discussed in the accompanying figure caption. Taking an RUC to describe the evolution operator $W(t)$ in these pictures, the next step is to evaluate the ensemble average (note that for simplicity we calculate $[e^{-S_2(t)}]_{\rm av}$ rather than $[S_2(t)]_{\rm av}$, which would require replica methods \cite{NahumEntanglementmembrane} that are beyond the level of these notes). The calculation requires a discussion of Haar averages, which we present next.

\begin{figure}[htb]
    \centering 
    \includegraphics[width=4cm]{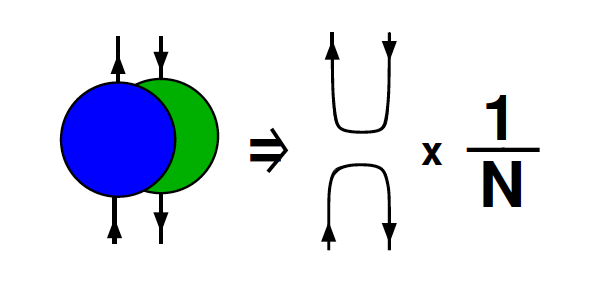}\hspace{1cm}
    \includegraphics[width=8cm]{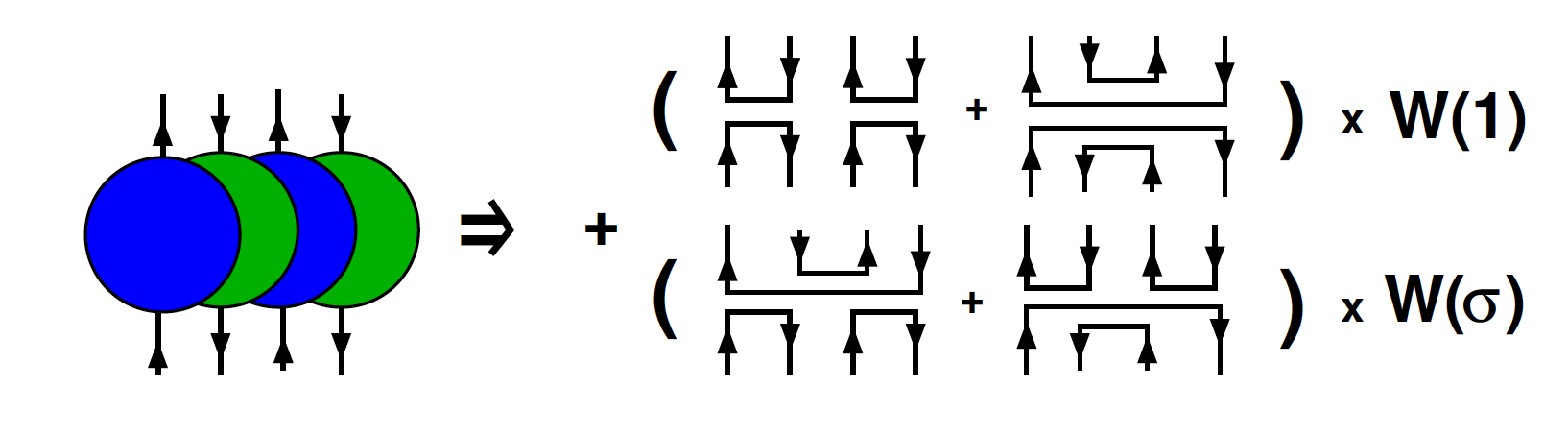}
    \caption{Pictorial representation of the two lowest-order Haar averages: the blue and green circles represent $U$ and $U^*$ respectively, and the right-hand sides of the expressions represent the Kronecker deltas in Eqns~\eqref{UU} and \eqref{UUUU}.
    }
    \label{fig:fig10}
\end{figure}

We consider the average over the Haar distribution of products of elements of an $N\times N$ unitary matrix $U$. The most general quantity of this kind has the form
\begin{equation}
    [U_{a_1\alpha_1} U_{a_2\alpha_2} \ldots U_{a_m \alpha_m} U^*_{b_1\beta_1}\ldots U^*_{b_n\beta_n}]_{\rm av},.
\end{equation}
A basic tool we can use to discuss such averages is the fact that the Haar distribution is invariant under $U\to UV_1$ and $U\to V_2U$ for any fixed unitary $V_1$, $V_2$. Taking $V_1$, $V_2$ to be diagonal, we see that the average is zero unless: (i) $m=n$; and (ii) $\{b_k\}$ is a permutation $\sigma$ of $\{a_k\}$;  and (iii) $\{\beta_k\}$ is a permutation $\tau$ of $\{\alpha_k \}$. The value of the average when these conditions are satisfied is known as a Weingarten function $W(\sigma^{-1}\tau)$ \cite{Creutz_1978,Samuel_1980,Brouwer_1996,Collins_2022}. These functions are relatively simple for general $N$ if $m$ is small, and also at large $N$ for general $m$. In particular, we have for the case $m=1$
\begin{equation}\label{UU}
    [U_{a_1\alpha_1} U^*_{b_1\beta_1}]_{\rm av} = \frac{1}{N} \delta_{a_1b_1}\delta_{\alpha_1\beta_1}\,.
\end{equation}
and for the case $m=2$
\begin{eqnarray}\label{UUUU}
    [U_{a_1\alpha_1}U_{a_2\alpha_2} U^*_{b_1\beta_1}U^*_{b_2\beta_2}]_{\rm av} 
    &=& W(1) [\delta_{a_1b_1}\delta_{a_2b_2}\delta_{\alpha_1\beta_1}\delta_{\alpha_2\beta_2}+ 
\delta_{a_1b_2}\delta_{a_2b_1}\delta_{\alpha_1\beta_2}\delta_{\alpha_2\beta_1}]\nonumber\\ 
&+& W(\sigma)[\delta_{a_1b_1}\delta_{a_2b_2}\delta_{\alpha_1\beta_2}\delta_{\alpha_2\beta_1}+ 
\delta_{a_1b_2}\delta_{a_2b_1}\delta_{\alpha_1\beta_1}\delta_{\alpha_2\beta_2}]\nonumber\\
\end{eqnarray}
with $W(1) = (N^2-1)^{-1}$ and $W(\sigma) = -[N(N^2-1)]^{-1}$.
We can represent these results pictorially as shown in Fig.~\ref{fig:fig10}. Note that the pairings shown on the right-hand sides of this figure ensure invariance under multiplication by the diagonal matrix $V$ discussed above.

\begin{figure}[htb]
    \centering 
    \includegraphics[width=5cm]{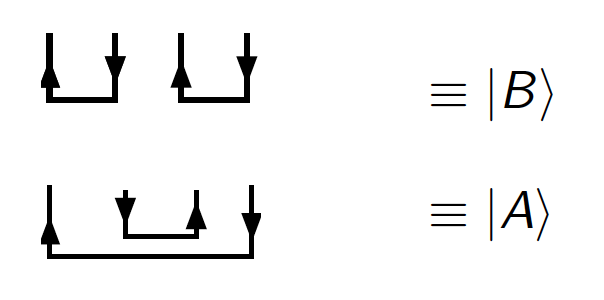}\hspace{2cm}
    \includegraphics[width=5cm]{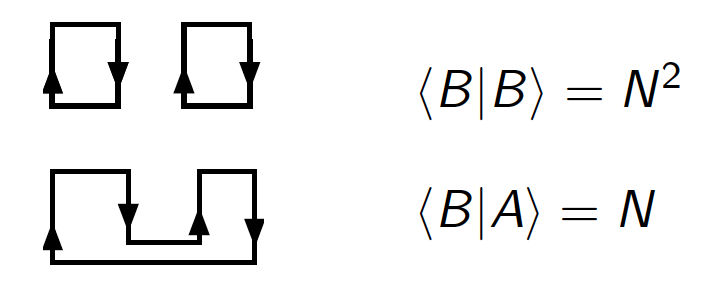}
    \caption{Left: vectors representing the two possible pairings at $m=2$. Right: evaluation of overlaps between these vectors.
    }
    \label{fig:fig11}
\end{figure}

To discuss the case $m=2$ it is useful to introduce the vector notation $|A\rangle$ and $|B\rangle$ for the two possible pairings of indices, as illustrated on the left side of Fig.~\ref{fig:fig11}. Note that these vectors are neither normalised nor orthogonal; their overlaps can be computed as illustrated on the right side of Fig.~\ref{fig:fig11}, by associating a factor of $N$ with each closed loop, arising from summation over the associated label.

Using this notation, we can represent Eq.~\eqref{UUUU} as 
\begin{equation}
    [UUU^*U^*]_{\rm av} \equiv \hat{M} = [|A\rangle\langle A| + |B\rangle\langle B| ]W(1) + [|A\rangle\langle B| + |B\rangle\langle A| ]W(\sigma) \,.
\end{equation}
Moreover, one finds 
\begin{equation}\label{MA}
\hat{M}|A\rangle = |A\rangle \quad \mbox{and} \quad \hat{M}|B\rangle = |B\rangle\,.
\end{equation}

We now apply these general results to the instance where $U$ is a two-site gate in the brickwork RUC introduced above, so that $N$ takes the value $q^2$. We note that the vectors $|A\rangle$ and $|B\rangle$ correspond to the pairings introduced in the respective subsystems in the calculation of purity that is illustrated in Fig.~\ref{fig:fig9}. We can therefore use Eq.~\eqref{MA} to turn the pictures in Fig~\ref{fig:fig9} into an algebraic expression. To complete this task, one detail remains for clarification. It arises because $\hat{M}$ acts on a pair of sites while the $A$ and $B$ pairings of Fig.~\ref{fig:fig9} are defined for individual sites and may be different at a pair of adjacent sites. This leads us to consider the action of $M$ on (in obvious notation) $|A_{\mbox{1-site}}\rangle\otimes |B_{\mbox{1-site}}\rangle$. It is straightforward to show that
\begin{equation}\label{eq:wt}
    M|A_{\mbox{1-site}}\rangle\otimes |B_{\mbox{1-site}}\rangle = \frac{q}{q^2+1}\big[|A\rangle + |B\rangle\big]\,.
\end{equation}
Finally, contraction of either $|A\rangle$ or $|B\rangle$ with the four copies of an initial state, appearing at the bottom in Fig.~\ref{fig:fig9}, yields unity when this is a product state in the site basis. 

\begin{figure}[htb]
    \centering 
    \includegraphics[width=4cm]{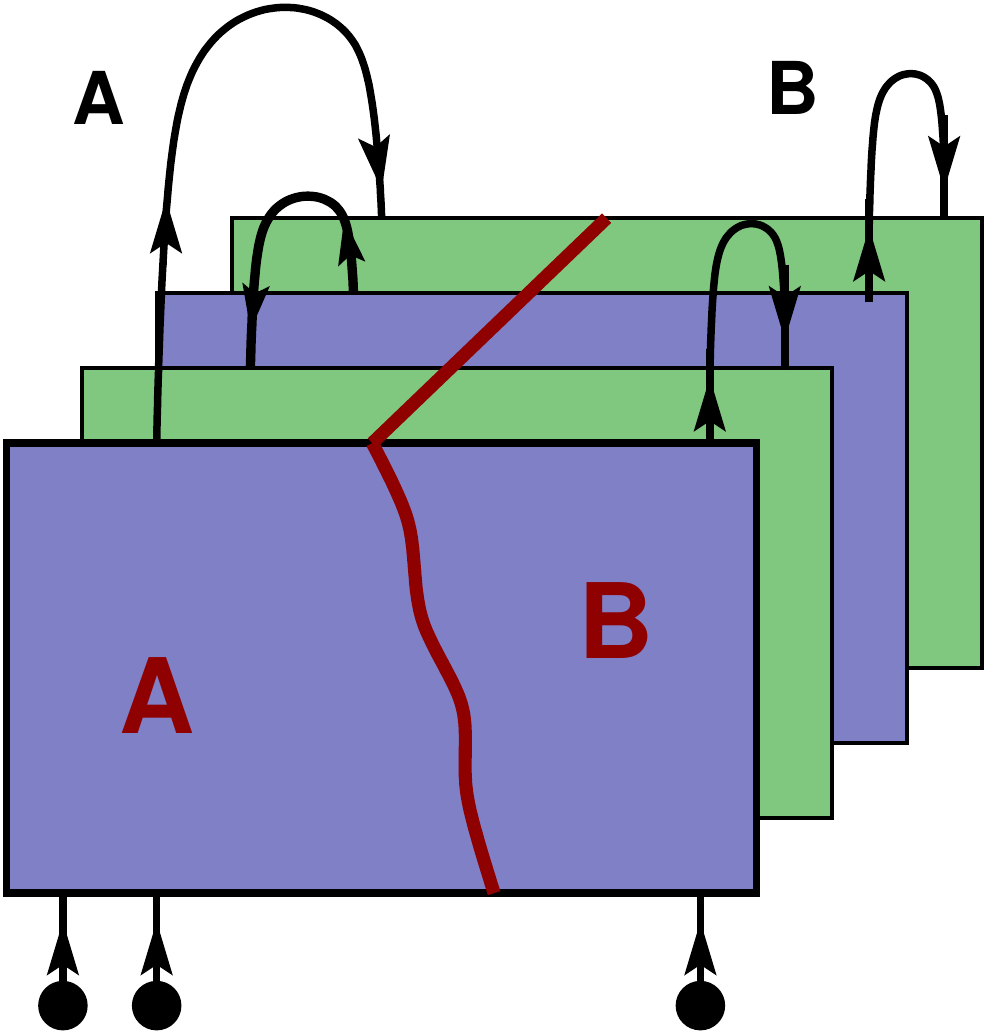}
    \hspace{2cm}
    \includegraphics[width=6cm]{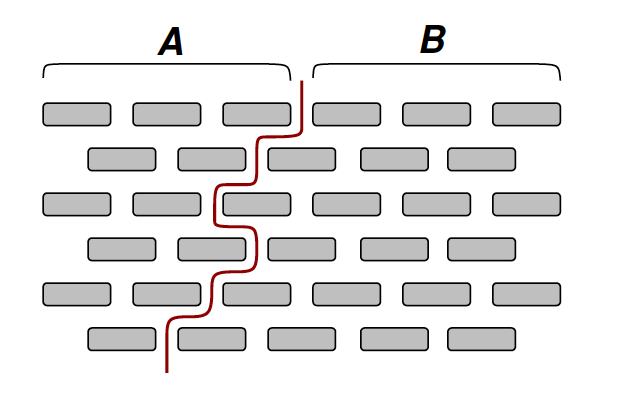}
    \caption{Directed random walk arising in a calculation of the average purity after time evolution by an RUC. Left: in the notation of Fig.~\ref{fig:fig9}. Right: showing the individual gates, with all four layers of the left-hand picture projected onto a single plane.
    }
    \label{fig:fig12}
\end{figure}

To summarise, the calculation shows that the average purity is given by a sum of contributions, each associated with a configuration of a directed random walk as illustrated in Fig.~\ref{fig:fig12}. From Eq.~\eqref{eq:wt} the walk has a weight $q/(q^2+1)$ per step. In the early-time regime there are two choices for each step of the walk, and so the purity is
$$
{\rm Tr}\, [\rho_A(t)^2]= \left(\frac{2q}{q^2+1}\right)^t\,.
$$
This random walk or domain wall between the two pairing states $A$ and $B$ is an example of an object that also appears in other calculations of entanglement, and is known as the entanglement membrane (in general, it is a $d$-dimensional hyper-surface for a system in $d$ space and one time dimension, so here it is one-dimensional) \cite{Ryu_2006,Jonay_2018,Mezei_2018}. Because the entanglement membrane has a tension (here, the weight per step), configurations of minimal area (here, minimal length) make the biggest contribution where they exist. This can result in a change in behaviour as a function of time. Suppose that the division between subsystems $A$ and $B$ lies deep within the system, with $1\ll L_A \ll L_B$. Then, as depicted in Fig.~\ref{fig:fig13b}, at short times the dominant paths end at the lower edge of the system, while at later times they  end at the edge of the system. This implies the dependence on time of second Renyi entropy $S_2(t)$ (ignoring possible differences between averages of $[S_2(t)]_{\rm av}$ and of $\ln [e^{-S_2(t)}]_{\rm av}$) that was derived in the \emph{min-cut} approach and is sketched in Fig.~\ref{fig:fig13a}. 
  
\section{Spectral correlations}

We now turn to the aspect of generic quantum systems that originally sparked the development of random matrix theory in the context of nuclear physics: the statistical properties of spectra. Rather than discuss Hamiltonian models as in that original work, we will deal with the time-evolution operator, and hence eigenvalues of unitary matrices, and in order to have a fixed evolution operator, we will consider Floquet systems, rather than circuits in which gates are chosen independently at each time-step. 

\begin{figure}[htb]
    \centering 
    \includegraphics[width=6cm]{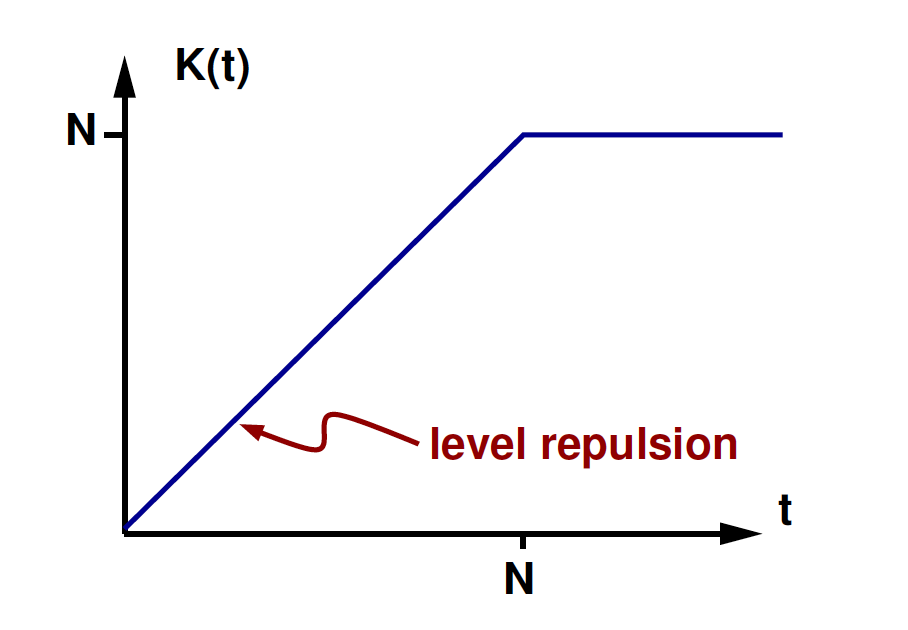}
    \caption{Behaviour of the spectral form factor $K(t)$ for $N\times N$ Haar-distributed random unitary matrices at times $t>0$
    }
    \label{fig:fig14}
\end{figure}

An $N\times N$ unitary matrix $W$ has $N$ eigenvalues $e^{i\theta_n}$, which all lie on the unit circle in the complex plane, and we are concerned with statistical properties of their distribution, particularly for $N$ large. A central quantity that characterises this distribution is the spectral form factor (SFF), which is the Fourier transform of the two-point correlation function of densities. Using the notation $W(t) \equiv W^t$ for integer time $t$, it is defined by
\begin{equation}\label{SFF}
    K(t) \equiv \big[|{\rm Tr}\,W(t)|^2\big]_{\rm av} = \Big[ \sum_{m n} e^{i(\theta_m - \theta_n)t}\Big]_{\rm av}\,.
\end{equation}
An important reference point is provided by the behaviour of the SFF in the case that $W$ is drawn from the Haar distribution, yielding \cite{Haake_2010}
\begin{equation}\label{SFFlargeN}
    K(t) = \left\{ \begin{array}{lll}N^2 &\quad & t=0\\ |t| && 1\leq |t| \leq N\\ N &&N \leq |t|
    \end{array}
    \right.
\end{equation}
as illustrated in Fig.~\ref{fig:fig14}. Some parts of this behaviour are immediately apparent from the right-hand expression in Eq.~\eqref{SFF}: at $t=0$ all $N^2$ terms in the double sum on $m$ and $n$ contribute unity, while at sufficiently large $t$ we can expect off-diagonal terms, $e^{i(\theta_m - \theta_n)t}$ for $m\not=n$, to average to zero, leaving contributions of unity from each of the $N$ diagonal terms. This simple plateau behaviour in fact extends down to the Heisenberg time $t=N$, a scale set by the inverse of the level spacing. The linear variation of $K(t)$ with $t$ in the time range below the Heisenberg time (known as the ramp) is a consequence of strong correlations between eigenvalues. Indeed, if the eigenvalues were Poisson distributed we would have $K(t) = N$ for all integer $t\not=0$: its suppression far below this value within the ramp reflects the small intensity of density fluctuations. 

\begin{figure}[htb]
    \centering 
    \includegraphics[width=14cm]{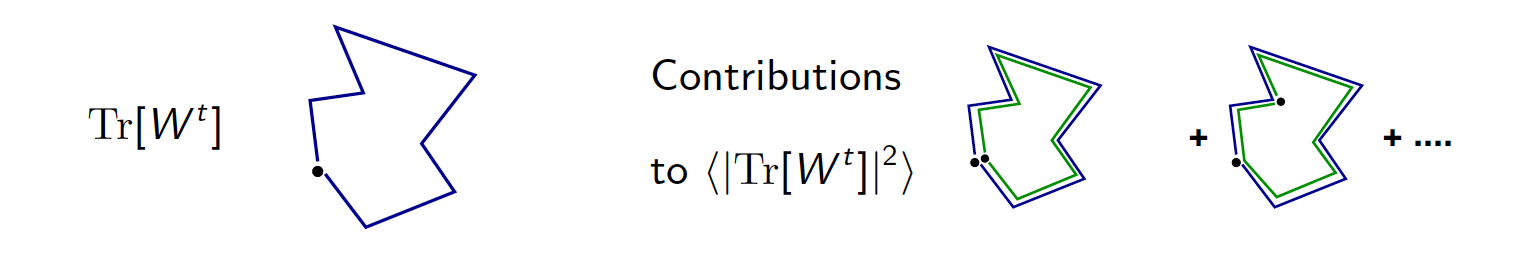}
    \caption{Expansion of the spectral form factor as a sum on Feynman paths through Hilbert space. 
    }
    \label{fig:fig15}
\end{figure}

Our overall interest in this section is in understanding spectral correlations in spatially extended system with local couplings. In order to introduce the techniques that we will use, we first discuss how the presence of the ramp in the SFF can be understood for Haar-distributed unitary matrices at large $N$, using the language of Feynman paths. To begin, note that ${\rm Tr}\,[W^t] =\sum_{a_1 \ldots a_t}W_{a_1a_2}W_{a_2a_3} \ldots W_{a_ta_1}$. We can think of each sequence $a_1 a_2 \ldots a_t$ as a closed path in Hilbert space and the product $W_{a_1a_2}W_{a_2a_3} \ldots W_{a_ta_1}$ as an associated Feynman amplitude. Similarly, ${\rm Tr}\,[W^t]^* =\sum_{b_1 \ldots b_t}W^*_{b_1b_2}W^*_{b_2b_3} \ldots W^*_{b_tb_1}$, so that $K(t)$ is given by a sum over pairs of paths, of the product of their amplitudes, in the form
$$
K(t) = \sum_{a_1 \ldots a_t} \sum_{b_1 \ldots b_t} \big[W_{a_1a_2} \ldots W_{a_ta_1} W^*_{b_1b_2}\ldots W_{b_tb_1}\big]_{\rm av}.
$$
In this sum we have constructive interference (in the form of positive contributions) from the terms in which the path $b_1b_2 \ldots b_t$ is a (possibly time-translated) copy of the path $a_1a_2 \ldots a_t$. Conversely, many of the terms that are not of this form have arbitrary phases, and it is reasonable to expect that they may average to zero in the ensemble. The key point now is that, at time $t$, there are $t$ such time-translated copies, and summing over these copies gives rise to the ramp in the SFF, as illustrated in Fig.~\ref{fig:fig15}.

\begin{figure}[htb]
    \centering 
    \includegraphics[width=12cm]{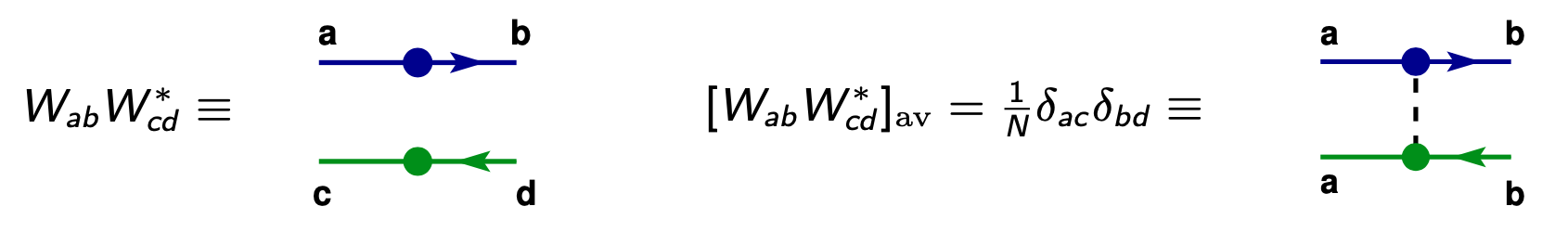}
    \caption{Ingredients for a diagrammatic calculation of the spectral form factor: (left) the unaveraged quantity $W_{ab}W^*_{cd}$; (right) the ensemble average $[W_{ab}W^*_{cd}]_{\rm av}$. 
    }
    \label{fig:fig16}
\end{figure}

It is useful and interesting to go beyond this qualitative argument, not simply deriving the relation $K(t) \propto t$, but also evaluating the constant of proportionality. This can be done for $N$ large, when Haar averages simplify because the elements $W_{ab}$ are then approximately Gaussian. This is expected because these elements are components of an orthonormal set of vectors, and the constraints of orthonormality are weak when $N$ is large; it is also evident at the leading two orders from Eqns.~\eqref{UU} and \eqref{UUUU}. Taking $W_{ab}$ to be Gaussian with zero mean and variance $[|W_{ab}|^2]_{\rm av} = N^{-1}$, the average can be computed using Wick's theorem by constructing all possible pairings of $W_{ab}$ with $W^*_{cd}$ and associating a factor of $N^{-1}\delta_{ac}\delta_{bd}$ with each. It is convenient to present this calculation diagrammatically, using the elements shown in Fig.~\ref{fig:fig16}.

\begin{figure}[htb]
    \centering 
    \includegraphics[width=13cm]{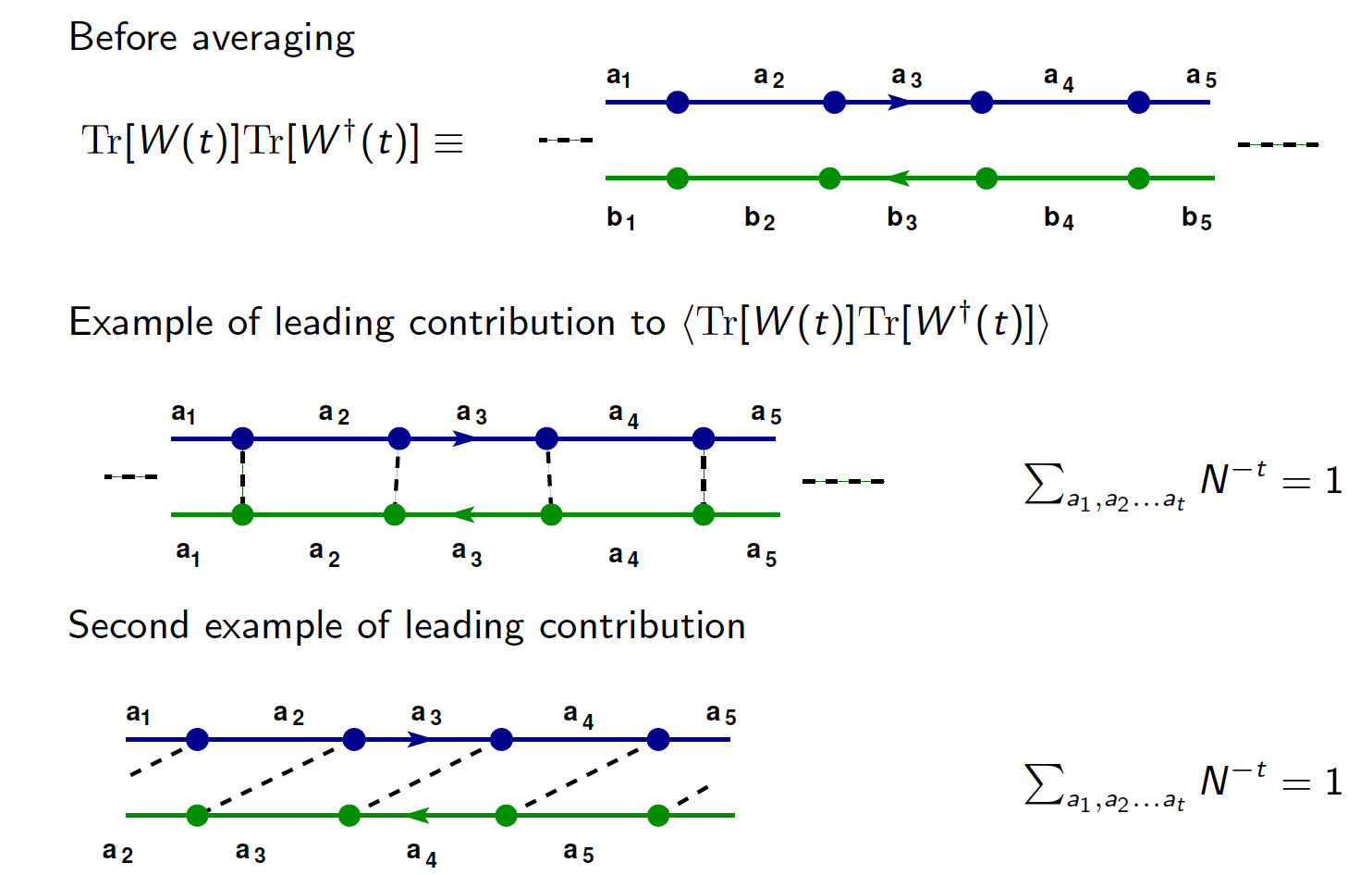}
    \caption{Evaluation of the spectral form factor for a Haar-distributed unitary matrix at large $N$. 
    }
    \label{fig:fig17}
\end{figure}

\begin{figure}[htb]
    \centering 
    \includegraphics[width=10cm]{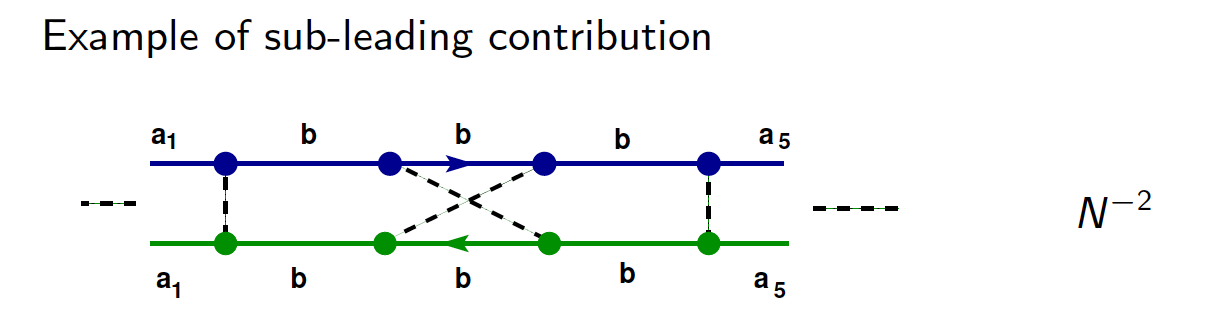}
    \caption{Example of a contribution to $K(t)$ that is sub-leading at large $N$.
    }
    \label{fig:fig18}
\end{figure}

The steps are shown in Fig.~\ref{fig:fig17}. Forming all Wick pairs of $W_{ab}$ with $W_{cd}^*$, there are $t$ leading order contributions to $K(t)$, two of which are depicted here. Each one has a weight $N^{-t}$ before summation over internal lables, from the averages of $t$ factors of $|W_{ab}|^2$ , and a weight of unity after this summation. Combining all of them them gives $K(t) =t$. The omitted terms are small in powers of $N^{-1}$, as illustrated in Fig.~\ref{fig:fig18}. In this way we recover the behaviour of the SFF given in Eq.~\eqref{SFFlargeN} up to the Heisenberg time. The plateau at later times does not appear from the large $N$ calculation we have outlined, and can only be obtained using quite different techniques \cite{Haake_2010,Efetov_1997}.

\begin{figure}[htb]
    \centering 
    \includegraphics[width=10cm]{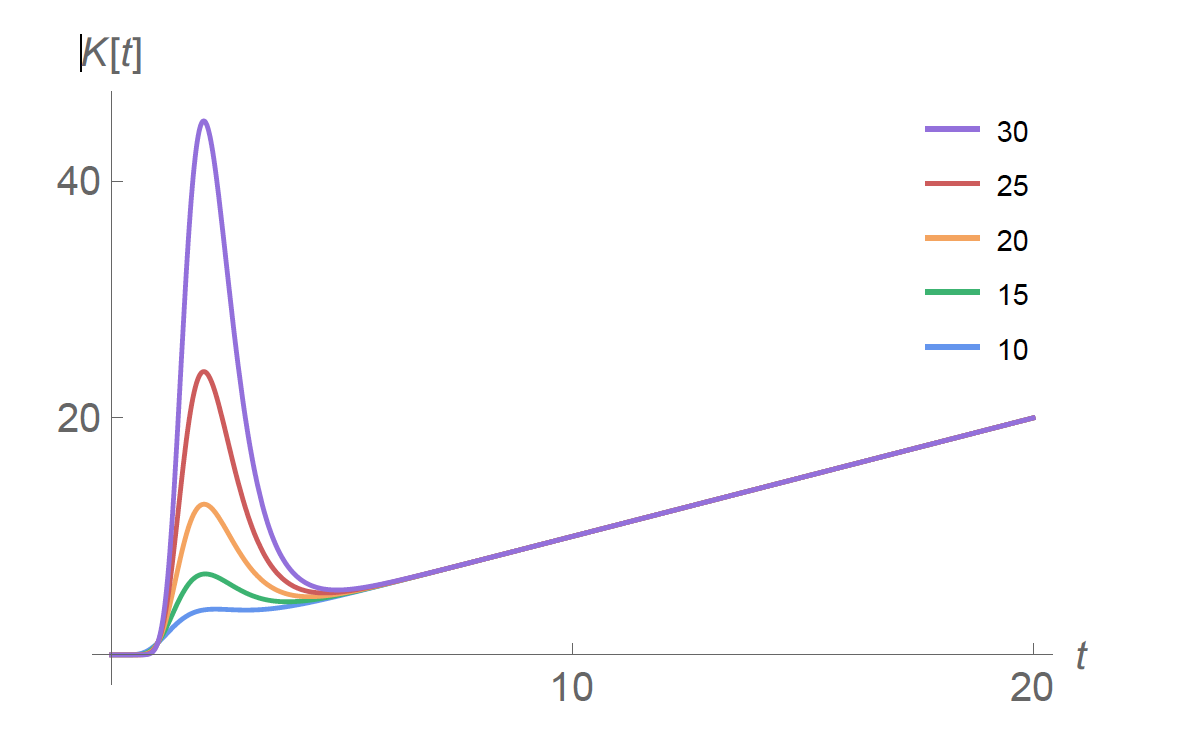}
    \caption{
    The spectral form factor $K(t)$ vs $t$ for the random phase model [Fig.~\ref{fig:fig3}] at the system sizes $L$ indicated, from Ref.~\cite{Chan_PRL}.
    }
    \label{fig:fig19}
\end{figure}

We now turn our attention from the reference problem of a single, Haar-distributed unitary matrix to spatially extended systems. As a concrete example, it is useful to have in mind the RPM, illustrated in Fig.~\ref{fig:fig3}. We will outline results obtained analytically for this model at large $q$ \cite{Chan_PRL}, which are in fact representative of the behaviour found more generally in numerical studies \cite{Garratt_2021}. As a starting  point, consider first the SFF for an uncoupled chain of sites, obtained by setting $\varepsilon=0$ in the RPM. In this simple limit, each site contributes a factor of $t$ to the SFF for $t\leq q$, and so in an $L$-site system the SFF grows very rapidly with $t$, as $K(t) = t^L$. Reintroducing weak intersite coupling by setting $\varepsilon>0$, we might expect similar behaviour at early times, before the coupling takes effect, but at sufficiently late times the idea of random matrix universality suggests a reversion to the form $K(t) =t$. This is indeed what is found \cite{Chan_PRL}, as illustrated in Fig.~\ref{fig:fig19}. 

\begin{figure}[htb]
    \centering 
    \includegraphics[width=10cm]{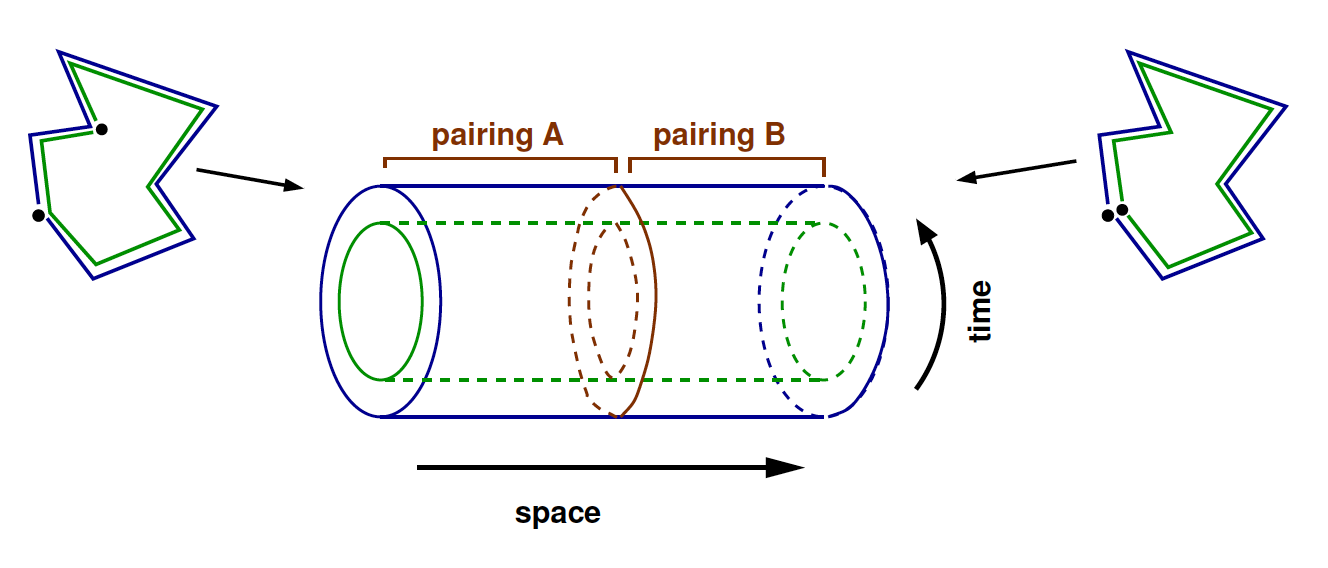}
    \caption{Contributions to the spectral form factor at early times in a spatially extended system: see main text for details.
    }
    \label{fig:fig20}
\end{figure}

This behaviour can be understood in terms of constructive interference between Feynman paths in Fock space \cite{Garratt_2021}, in an extension of the ideas sketched in Fig.~\ref{fig:fig15}. To discuss Feynman paths contributing to ${\rm Tr}\, W(t)$ in a one-dimensional, spatially extended system, we should consider a cylinder, with the space coordinate running along the length of the cylinder and the time-coordinate wrapping around the cylinder, and to represent $|{\rm Tr}\, W(t)|^2$ we are led to think about two concentric cylinders. Two Feynman paths are specified by giving values for the states of the system at each space-time point on the two cylinders. By the same arguments that have been presented for a single Haar matrix, we again expect constructive interference between paired paths, in which the labels on one path are the same as those on the other path, up to a possible translation in time, as illustrated in Fig.~\ref{fig:fig15}. However, in a spatially extended system with local couplings, a new possibility arises, which is that the nature of this pairing (that is, the relative time shift of the path in ${\rm Tr} \, [W^t]$ relative to the one in ${\rm Tr} \, [W^t]^*$) varies as a function of position. This possibility leads to domains of different pairing, separated by domain walls, as illustrated in Fig.~\ref{fig:fig20}. These domain walls carry a statistical penalty, which in the RPM at large $q$ is $e^{-\varepsilon t}$ \cite{Chan_PRL}. 

The resulting form for the SFF in an $L$-site system is (taking open boundary conditions for simplicity)
\begin{equation}\label{eq:K}
    K(t) = t[1+(t-1)e^{-\varepsilon t}]^{L-1}\,.
\end{equation}
The expression has a straightforward interpretation. Working from one end of the system, the leading factor of $t$ arises because this is the number of possible pairings for the first site. Each further site may either carry the same pairing, or one of $(t-1)$ different pairings, and the weights for both alternatives are combined in the factor $[1+(t-1)e^{-\varepsilon t}]$. Expansion of the right-hand side of Eq.~\eqref{eq:K} therefore generates a sum of terms corresponding to all possible pairings at each site of the system, with domain walls between sites that carry different pairings. This function is illustrated in Fig.~\ref{fig:fig19}.

At early times $K(t)$ is greatly enhanced compared to the random matrix form, with a timescale $t_{\rm Th}$ for crossover to the latter known that is known as the Thouless time. This name for the timescale of crossover to random matrix behaviour is used widely, although the underlying physics and hence the dependence of the Thouless time on system size varies according to context. It was first introduced in discussions of single-particle models of disordered conductors \cite{Thouless_77} and in that setting it is controlled by diffusion of probability density, varying with diffusion constant $D$ and system size $L$ as $t_{\rm Th}\sim L^2/D$. By contrast, in the many-body setting of Floquet circuits, it is the timescale at which domain walls between different pairings (discussed above) are suppressed: from Eq.~\eqref{eq:K} this varies with system size as $t_{\rm Th} \sim\ln L$. 

We note that the domain walls appearing here are similar in spirit but different in detail to the entanglement membrane discussed in Sec.~\ref{sec:entanglement}. The similarities are that in both cases we are picking out from a sum over multiple sets of Feynman paths some contributions that result in constructive interference; these involve pairing the paths, and there are multiple ways to arrange the pairing, with a domain wall or entanglement membrane as the interface between different pairings. The differences between the two cases are, first, that the purity involves two factors of $W(t)$ and two of $W^\dagger(t)$, while the SFF involves only one of each, and second, that boundary conditions in the time direction are different in the two cases. 

\section{Concluding remarks}

In summary, we have seen that a powerful route to understanding the physical properties of generic quantum systems is to construct an ensemble of macroscopically similar realisations, and to evaluate averages of informative quantities over the ensemble. The attraction of this route is partly that it can often be followed in a straightforward way. Appropriately enough for a summer school that includes fundamental problems from statistical physics more broadly, the simplifications one arrives at by this means often involve mappings to models from classical statistical physics. 

There are many closely related topics that we have not touched on in these notes. These include dual-unitary circuits \cite{Goplalkrishnan_2019,Bertini_2019,Bertini_2025}, for which many exact results can be obtained; circuits with Clifford gates \cite{Nielson_2000}, for which efficient classical simulations are possible, and fermionic systems \cite{Bauer_2019}. More broadly, a  complementary framework to the microscopic models of dynamics that we have discussed is provided by the eigenstate thermalisation hypothesis \cite{DAlessio_2016}.

\section*{Acknowledgements}

This work was supported in part by the EPSRC under grant EP/X030881/1. I am very grateful for discussions with Adam Nahum, and with my collaborators in this area: Amos Chan, Andrea De Luca, Aaron Friedman, Sam Garratt, Sthitadhi Roy, David Luitz and Dominik Hahn. I would also like to thank Oliver Breach for very helpful comments on a draft of these notes. 


\bibliography{references2}

@BOOK{Mehta_Random_2004,
  title     = "Random Matrices",
  author    = "Mehta, M L",
  publisher = "Academic Press",
  series    = "Pure and Applied Mathematics",
  edition   =  3,
  month     =  nov,
  year      =  2004
}

@article{DAlessio_2016,
author = {Luca D'Alessio and Yariv Kafri and Anatoli Polkovnikov and Marcos Rigol},
title = {From quantum chaos and eigenstate thermalization to statistical mechanics and thermodynamics},
journal = {Adv. Phys.},
volume = {65},
number = {3},
pages = {239-362},
year  = {2016},
publisher = {Taylor & Francis},
doi = {10.1080/00018732.2016.1198134},
URL = {https://doi.org/10.1080/00018732.2016.1198134}
}

@article{Fisher_Random_2023,
author = {Fisher, Matthew P. A. and Khemani, Vedika and Nahum, Adam and Vijay, Sagar},
title = {Random Quantum Circuits},
journal = {Annu. Rev. Condens. Matter Phys.},
volume = {14},
number = {1},
pages = {335-379},
year = {2023},
doi = {10.1146/annurev-conmatphys-031720-030658},
URL = {https://doi.org/10.1146/annurev-conmatphys-031720-030658}
}

@article{Thouless_77,
  title = {Maximum Metallic Resistance in Thin Wires},
  author = {Thouless, D. J.},
  journal = {Phys. Rev. Lett.},
  volume = {39},
  issue = {18},
  pages = {1167--1169},
  numpages = {0},
  year = {1977},
  month = {Oct},
  publisher = {American Physical Society},
  doi = {10.1103/PhysRevLett.39.1167},
  url = {https://link.aps.org/doi/10.1103/PhysRevLett.39.1167}
}

@article{Chan_PRX,
  title = {Solution of a Minimal Model for Many-Body Quantum Chaos},
  author = {Chan, Amos and De Luca, Andrea and Chalker, J. T.},
  journal = {Phys. Rev. X},
  volume = {8},
  issue = {4},
  pages = {041019},
  numpages = {17},
  year = {2018},
  month = {Nov},
  publisher = {American Physical Society},
  doi = {10.1103/PhysRevX.8.041019},
  url = {https://link.aps.org/doi/10.1103/PhysRevX.8.041019}
}

@article{Garratt_2021,
  title = {Local Pairing of {F}eynman Histories in Many-Body {F}loquet Models},
  author = {Garratt, S. J. and Chalker, J. T.},
  journal = {Phys. Rev. X},
  volume = {11},
  issue = {2},
  pages = {021051},
  numpages = {31},
  year = {2021},
  month = {Jun},
  publisher = {American Physical Society},
  doi = {10.1103/PhysRevX.11.021051},
  url = {https://link.aps.org/doi/10.1103/PhysRevX.11.021051}
}

@article{Chan_PRL,
  title = {Spectral Statistics in Spatially Extended Chaotic Quantum Many-Body Systems},
  author = {Chan, Amos and De Luca, Andrea and Chalker, J. T.},
  journal = {Phys. Rev. Lett.},
  volume = {121},
  issue = {6},
  pages = {060601},
  numpages = {5},
  year = {2018},
  month = {Aug},
  publisher = {American Physical Society},
  doi = {10.1103/PhysRevLett.121.060601},
  url = {https://link.aps.org/doi/10.1103/PhysRevLett.121.060601}
}

@BOOK{Efetov_1997,
  title     = "Supersymmetry in Disorder and Chaos",
  author    = "Efetov, K.",
  publisher = "Cambridge University Press",
  year      =  "1997"
}

@article{NahumQuantumentanglement,
  title = {Quantum Entanglement Growth under Random Unitary Dynamics},
  author = {Nahum, Adam and Ruhman, Jonathan and Vijay, Sagar and Haah, Jeongwan},
  journal = {Phys. Rev. X},
  volume = {7},
  issue = {3},
  pages = {031016},
  numpages = {30},
  year = {2017},
  month = {Jul},
  publisher = {American Physical Society},
  doi = {10.1103/PhysRevX.7.031016},
  url = {https://link.aps.org/doi/10.1103/PhysRevX.7.031016}
}

@article{NahumEntanglementmembrane,
  title = {Entanglement Membrane in Chaotic Many-Body Systems},
  author = {Zhou, Tianci and Nahum, Adam},
  journal = {Phys. Rev. X},
  volume = {10},
  issue = {3},
  pages = {031066},
  numpages = {37},
  year = {2020},
  month = {Sep},
  publisher = {American Physical Society},
  doi = {10.1103/PhysRevX.10.031066},
  url = {https://link.aps.org/doi/10.1103/PhysRevX.10.031066}
}

@BOOK{Haake_2010,
  title     = "Quantum Signature of Chaos",
  author    = "Haake, F",
  publisher = "Springer",
  series    = "Springer Series in Synergetics",
  edition   =  3,
  year      =  2010,
}

@article{Lieb_1972,
  title = {The finite group velocity of quantum spin systems},
  author = {Lieb, E. H. and Robinson, D. W.},
  journal = {Comm. Math. Phys.},
  volume = {28},
  pages = {251},
  year = {1972},
  month = {Sep},
  publisher = {Springer},
  doi = {10.1007/BF01645779},
  url = {https://link.springer.com/article/10.1007/BF01645779}
}

@Inbook{Potter_2022,
author="Potter, Andrew C.
and Vasseur, Romain",
editor="Bayat, Abolfazl
and Bose, Sougato
and Johannesson, Henrik",
title="Entanglement Dynamics in Hybrid Quantum Circuits",
bookTitle="Entanglement in Spin Chains: From Theory to Quantum Technology Applications",
year="2022",
publisher="Springer International Publishing",
address="Cham, Switzerland",
pages="211--249",
isbn="978-3-031-03998-0",
doi="10.1007/978-3-031-03998-0_9",
url="https://doi.org/10.1007/978-3-031-03998-0_9"
}

@article{NahumSpreading,
  title = {Operator Spreading in Random Unitary Circuits},
  author = {Nahum, Adam and Vijay, Sagar and Haah, Jeongwan},
  journal = {Phys. Rev. X},
  volume = {8},
  issue = {2},
  pages = {021014},
  numpages = {30},
  year = {2018},
  month = {Apr},
  publisher = {American Physical Society},
  doi = {10.1103/PhysRevX.8.021014},
  url = {https://link.aps.org/doi/10.1103/PhysRevX.8.021014}
}

@article{vonKeyserlingk_2018,
  title = {Operator Hydrodynamics, {OTOC}s, and Entanglement Growth in Systems without Conservation Laws},
  author = {von Keyserlingk, C. W. and Rakovszky, Tibor and Pollmann, Frank and Sondhi, S. L.},
  journal = {Phys. Rev. X},
  volume = {8},
  issue = {2},
  pages = {021013},
  numpages = {19},
  year = {2018},
  month = {Apr},
  publisher = {American Physical Society},
  doi = {10.1103/PhysRevX.8.021013},
  url = {https://link.aps.org/doi/10.1103/PhysRevX.8.021013}
}

@article{Brody_1981,
  title = {Random-matrix physics: spectrum and strength fluctuations},
  author = {Brody, T. A. and Flores, J. and French, J. B. and Mello, P. A. and Pandey, A. and Wong, S. S. M.},
  journal = {Rev. Mod. Phys.},
  volume = {53},
  issue = {3},
  pages = {385--479},
  numpages = {0},
  year = {1981},
  month = {Jul},
  publisher = {American Physical Society},
  doi = {10.1103/RevModPhys.53.385},
  url = {https://link.aps.org/doi/10.1103/RevModPhys.53.385}
}

@article{Dyson_1962,
    author = {Dyson, Freeman J.},
    title = {Statistical Theory of the Energy Levels of Complex Systems. {I}},
    journal = {Journal of Mathematical Physics},
    volume = {3},
    number = {1},
    pages = {140-156},
    year = {1962},
    month = {01},
    issn = {0022-2488},
    doi = {10.1063/1.1703773},
    url = {https://doi.org/10.1063/1.1703773},
}

@misc{Chalker_2025,
      title={Chaotic many-body quantum dynamics, spectral correlations, and energy diffusion}, 
      author={J. T. Chalker and Dominik Hahn},
      year={2025},
      eprint={2510.02198},
      archivePrefix={arXiv},
      primaryClass={quant-ph},
      url={https://arxiv.org/abs/2510.02198}, 
}

@article{Rakovsky_2018,
  title = {Diffusive Hydrodynamics of Out-of-Time-Ordered Correlators with Charge Conservation},
  author = {Rakovszky, Tibor and Pollmann, Frank and von Keyserlingk, C. W.},
  journal = {Phys. Rev. X},
  volume = {8},
  issue = {3},
  pages = {031058},
  numpages = {28},
  year = {2018},
  month = {Sep},
  publisher = {American Physical Society},
  doi = {10.1103/PhysRevX.8.031058},
  url = {https://link.aps.org/doi/10.1103/PhysRevX.8.031058}
}

@article{Khemani_2018,
  title = {Operator Spreading and the Emergence of Dissipative Hydrodynamics under Unitary Evolution with Conservation Laws},
  author = {Khemani, Vedika and Vishwanath, Ashvin and Huse, David A.},
  journal = {Phys. Rev. X},
  volume = {8},
  issue = {3},
  pages = {031057},
  numpages = {25},
  year = {2018},
  month = {Sep},
  publisher = {American Physical Society},
  doi = {10.1103/PhysRevX.8.031057},
  url = {https://link.aps.org/doi/10.1103/PhysRevX.8.031057}
}

@article{Yoshimura_2025,
  title = {Operator dynamics in Floquet many-body systems},
  author = {Yoshimura, Takato and Garratt, Samuel J. and Chalker, J. T.},
  journal = {Phys. Rev. B},
  volume = {111},
  issue = {9},
  pages = {094316},
  numpages = {28},
  year = {2025},
  month = {Mar},
  publisher = {American Physical Society},
  doi = {10.1103/PhysRevB.111.094316},
  url = {https://link.aps.org/doi/10.1103/PhysRevB.111.094316}
}

@article{Brouwer_1996,
    author = {Brouwer, P. W. and Beenakker, C. W. J.},
    title = {Diagrammatic method of integration over the unitary group, with applications to quantum transport in mesoscopic systems},
    journal = {Journal of Mathematical Physics},
    volume = {37},
    number = {10},
    pages = {4904-4934},
    year = {1996},
    month = {10},
    issn = {0022-2488},
    doi = {10.1063/1.531667},
    url = {https://doi.org/10.1063/1.531667},
}

@article{Creutz_1978,
    author = {Creutz, Michael},
    title = {On invariant integration over {SU(N)}},
    journal = {Journal of Mathematical Physics},
    volume = {19},
    number = {10},
    pages = {2043-2046},
    year = {1978},
    month = {10},
    issn = {0022-2488},
    doi = {10.1063/1.523581},
    url = {https://doi.org/10.1063/1.523581},
}

@article{Samuel_1980,
    author = {Samuel, Stuart},
    title = {{U(N)} Integrals, {1/N}, and the {De Wit}–{’t Hooft} anomalies},
    journal = {Journal of Mathematical Physics},
    volume = {21},
    number = {12},
    pages = {2695-2703},
    year = {1980},
    month = {12},
    issn = {0022-2488},
    doi = {10.1063/1.524386},
    url = {https://doi.org/10.1063/1.524386},
}

@article{Collins_2022,
   title={The {W}eingarten Calculus},
   volume={69},
   ISSN={1088-9477},
   url={http://dx.doi.org/10.1090/noti2474},
   DOI={10.1090/noti2474},
   number={05},
   journal={Notices of the American Mathematical Society},
   publisher={American Mathematical Society (AMS)},
   author={Collins, Benoit and Matsumoto, Sho and Novak, Jonathan},
   year={2022},
   month=may, pages={1} }

@BOOK{Nielson_2000,
  title     = "Quantum computation and quantum information",
  author    = "Nielson, M. A. and Chuang, I. L.",
  publisher = "Cambridge University Press Press",
  year      =  2000
}

@misc{Jonay_2018,
      title={Coarse-grained dynamics of operator and state entanglement}, 
      author={Cheryne Jonay and David A. Huse and Adam Nahum},
      year={2018},
      eprint={1803.00089},
      archivePrefix={arXiv},
      primaryClass={cond-mat.stat-mech},
      url={https://arxiv.org/abs/1803.00089}, 
}

@article{Casini_2016,
   title={Spread of entanglement and causality},
   volume={77},
   url={https://doi.org/10.1007/JHEP07(2016)077},
   DOI={10.1090/noti2474},
   journal={Journal of High Energy Physics},
   publisher={Springer},
   author={Casini, Horacio and Liu, Hong and Mezei, M\a'rk},
   year={2016},
   }

@article{Swingle_2012,
  title = {Entanglement renormalization and holography},
  author = {Swingle, Brian},
  journal = {Phys. Rev. D},
  volume = {86},
  issue = {6},
  pages = {065007},
  numpages = {8},
  year = {2012},
  month = {Sep},
  publisher = {American Physical Society},
  doi = {10.1103/PhysRevD.86.065007},
  url = {https://link.aps.org/doi/10.1103/PhysRevD.86.065007}
}

@article{Mezei_2018,
  title = {Membrane theory of entanglement dynamics from holography},
  author = {Mezei, M\'ark},
  journal = {Phys. Rev. D},
  volume = {98},
  issue = {10},
  pages = {106025},
  numpages = {9},
  year = {2018},
  month = {Nov},
  publisher = {American Physical Society},
  doi = {10.1103/PhysRevD.98.106025},
  url = {https://link.aps.org/doi/10.1103/PhysRevD.98.106025}
}

@article{Bertini_2019,
  title = {Exact Correlation Functions for Dual-Unitary Lattice Models in $1+1$ Dimensions},
  author = {Bertini, Bruno and Kos, Pavel and Prosen, Tomaz},
  journal = {Phys. Rev. Lett.},
  volume = {123},
  issue = {21},
  pages = {210601},
  numpages = {6},
  year = {2019},
  month = {Nov},
  publisher = {American Physical Society},
  doi = {10.1103/PhysRevLett.123.210601},
  url = {https://link.aps.org/doi/10.1103/PhysRevLett.123.210601}
}

@misc{Bertini_2025,
      title={Exactly solvable many-body dynamics from space-time duality}, 
      author={Bruno, Bertini and Claeys, Pieter W. and Tomaž, Prosen},
      year={2025},
      eprint={2505.11489},
      archivePrefix={arXiv},
      primaryClass={cond-mat.stat-mech},
      url={https://arxiv.org/abs/2505.11489}
}

@article{Bauer_2019,
   title={Equilibrium fluctuations in maximally noisy extended quantum systems},
   volume={6},
   ISSN={2542-4653},
   url={http://dx.doi.org/10.21468/SciPostPhys.6.4.045},
   DOI={10.21468/scipostphys.6.4.045},
   number={4},
   journal={SciPost Physics},
   publisher={Stichting SciPost},
   author={Bauer, Michel and Bernard, Denis and Jin, Tony},
   year={2019},
   month=apr }

@article{Goplalkrishnan_2019,
  title = {Unitary circuits of finite depth and infinite width from quantum channels},
  author = {Gopalakrishnan, Sarang and Lamacraft, Austen},
  journal = {Phys. Rev. B},
  volume = {100},
  issue = {6},
  pages = {064309},
  numpages = {15},
  year = {2019},
  month = {Aug},
  publisher = {American Physical Society},
  doi = {10.1103/PhysRevB.100.064309},
  url = {https://link.aps.org/doi/10.1103/PhysRevB.100.064309}
}

@article{Ryu_2006,
  title = {Holographic Derivation of Entanglement Entropy from the anti--{de Sitter Space/Conformal Field Theory Correspondence}},
  author = {Ryu, Shinsei and Takayanagi, Tadashi},
  journal = {Phys. Rev. Lett.},
  volume = {96},
  issue = {18},
  pages = {181602},
  numpages = {4},
  year = {2006},
  month = {May},
  publisher = {American Physical Society},
  doi = {10.1103/PhysRevLett.96.181602},
  url = {https://link.aps.org/doi/10.1103/PhysRevLett.96.181602}
}
\end{document}